\documentclass[a4paper,10pt]{article}

\usepackage[utf8]{inputenc}

\usepackage{authblk}
\usepackage{textcomp}
\usepackage{ifthen,ifpdf}
\usepackage[utf8]{inputenc}
\usepackage[T1]{fontenc}

\usepackage{breqn}
\usepackage{lscape}
\usepackage{scalerel,stackengine}
\usepackage{amsmath, amsthm, amssymb}
\usepackage{bm}
\usepackage{mathtools}
\usepackage{enumerate}
\usepackage{enumitem}
\usepackage{afterpage}
\usepackage{tabularx}
\usepackage{dsfont}
\usepackage{physics}
\usepackage{ulem} %
\usepackage{graphicx}
\usepackage[dvipsnames]{xcolor}

\usepackage{algorithm}
\usepackage{algpseudocode}
\usepackage{caption}
\usepackage[labelformat=simple]{subcaption}

\usepackage{color}
\usepackage{units}
\usepackage{array, multirow}
\usepackage{booktabs}
\newcolumntype{M}[1]{>{\vspace{3pt}\raggedleft\arraybackslash}m{#1}}

\usepackage{tikz}
\usepackage{pgfplots}
\usepackage{pgfplotstable}
\usetikzlibrary{angles, quotes, calc, shapes.geometric, positioning, spy} 
\usepgfplotslibrary{units}
\pgfplotsset{compat=1.9}
\pgfplotstableset{col sep=comma}

\usepackage{cite}
\usepackage{anysize} 
\marginsize{2.5cm}{2.5cm}{2cm}{2cm}
\usepackage{wrapfig}

\usepackage{pstool}
\usepackage{diagbox}
\usepackage{import}
\usepackage{ stmaryrd }
\SetSymbolFont{stmry}{bold}{U}{stmry}{m}{n}
\usepackage{hyperref}
\hypersetup{colorlinks=false,allcolors=blue}
\usepackage{hypcap}
\pdfpagewidth=\paperwidth
\pdfpageheight=\paperheight

\theoremstyle{plain}

\theoremstyle{remark}

\numberwithin{equation}{section}

\newcommand{\fonems}[1]{#1} %
\newcommand{\fonetb}[1]{#1} 
\newcommand{\foneme}[1]{#1}

\newcommand{\ftwoms}[1]{#1} %
\newcommand{\ftwotb}[1]{#1}
\newcommand{\ftwome}[1]{#1}

\newcommand{\fthrms}[1]{#1} %
\newcommand{\fthrtb}[1]{#1}
\newcommand{\fthrme}[1]{#1}

\newcommand{\ffoutb}[1]{#1} %
\newcommand{\ffouha}[1]{#1} %

\newcommand{\mysf}[1]{#1}

\newcommand{\sty}[1]{\mbox{\boldmath $#1$}}
\newcommand{\styy}[1]{{\mathbb{#1}}}

\newcommand{\fa}{\sty{ a}}
\newcommand{\fb}{\sty{ b}}

\newcommand{\fd}{\sty{ d}}
\newcommand{\fe}{\sty{ e}}

\newcommand{\fg}{\sty{ g}}

\newcommand{\fm}{\sty{ m}}
\newcommand{\fn}{\sty{ n}}

\newcommand{\fp}{\sty{ p}}
\newcommand{\fq}{\sty{ q}}

\newcommand{\fu}{\sty{ u}}
\newcommand{\fv}{\sty{ v}}
\newcommand{\fw}{\sty{ w}}
\newcommand{\fx}{\sty{ x}}

\newcommand{\fA}{\sty{ A}}
\newcommand{\fB}{\sty{ B}}

\newcommand{\fD}{\sty{ D}}

\newcommand{\fI}{\sty{ I}}

\newcommand{\fL}{\sty{ L}}

\newcommand{\fN}{\sty{ N}}

\newcommand{\fQ}{\sty{ Q}}
\newcommand{\fR}{\sty{ R}}

\newcommand{\fX}{\sty{ X}}
\newcommand{\fY}{\sty{ Y}}

\newcommand{\ffA}{\styy{ A}}

\newcommand{\ffC}{\styy{ C}}

\newcommand{\ffI}{\styy{ I}}

\newcommand{\ffK}{\styy{ K}}

\newcommand{\ffM}{\styy{ M}}

\newcommand{\ffP}{\styy{ P}}

\newcommand{\ffR}{\styy{ R}}

\newcommand{\ffT}{\styy{ T}}

\newcommand{\ffV}{\styy{ V}}

\newcommand{\ffX}{\styy{ X}}

\newcommand{\ftau}{\mbox{\boldmath $\tau$}}
\newcommand{\fsigma}{\mbox{\boldmath $\sigma$}}

\newcommand{\gdot}{\dot{\gamma}}
\newcommand{\persec}{\, \unit{s}^{-1}}
\newcommand{\Pas}{\unit{Pa} \, \unit{s}}

\newcommand{\fvhat}{\hat{\fv}}

\newcommand{\ra}{r_{\sf{a}}}
\newcommand{\cf}{c_{\sf{F}}}
\newcommand{\T}[1]{{#1}^{\sf{T}}}

\newcommand{\dif}{\mathop{}\!\mathrm{d}}

\newcommand{\solid}[1]{#1^{\sf{co}}}
\newcommand{\fluid}[1]{#1}
\newcommand{\ico}[1]{#1^{\sf{ico}}}
\newcommand{\spaceAstrain}{{\mathcal{E}}^{\sf{A}}}
\newcommand{\spaceAstrainsfr}{{\mathcal{E}}^{\sf{A}}_{\mysf{r}}}
\newcommand{\spacestressco}{\solid{\mathcal{S}}}
\newcommand{\spacestrainco}{\solid{\mathcal{E}}}
\newcommand{\spacestressico}{\ico{\mathcal{S}}}
\newcommand{\spacestrainico}{\ico{\mathcal{E}}}

\newcommand{\Yfluid}{\fluid{Y}}

\newcommand{\GammaA}{\ffGamma_{\fthrms{\ffI}}}
\newcommand{\GammaAtwo}{\ffGamma_{\ffA_2}}
\newcommand{\GammaAzero}{\ffGamma_{\ffA_0}}

\newcommand{\Gammasolidone}{\ffGamma_1^{\sf{co}}}
\newcommand{\Gammasolidtwo}{\ffGamma_2^{\sf{co}}}
\newcommand{\Gammafluidone}{\ffGamma_1^{\sf{ico}}}
\newcommand{\Gammafluidtwo}{\ffGamma_2^{\sf{ico}}}

\newcommand{\dpotk}{\Psi_{\mysf{k}}}

\newcommand{\Deff}{\bar{\fD}}
\newcommand{\Dload}{\tilde{\fD}}

\newcommand{\Aeff}{\bar{\ffA}}
\newcommand{\Veff}{\bar{\ffV}}
\newcommand{\Vsample}{\ffV^{\mysf{s}}}

\newcommand{\Veffsample}{\bar{\ffV}^{\mysf{s}}}
\newcommand{\Meff}{\bar{\ffM}}
\newcommand{\Keff}{\bar{\ffK}}
\newcommand{\Leff}{\bar{\fL}}
\newcommand{\geff}{\bar{\fg}}

\newcommand{\sigmaeff}{\bar{\fsigma}}

\newcommand{\sym}{\sf{Sym}(3)}
\newcommand{\symdev}{{\sf{Sym}_0(3)}}

\newcommand{\symfour}{\sf{M}}
\newcommand{\symdevfour}{\sf{M}_0}
\newcommand{\symplusfour}{\sf{M}^{+}}
\newcommand{\symdevplusfour}{\sf{M}_0^{+}}
\newcommand{\vmean}[1]{\left\langle {#1} \right\rangle_Y}

\newcommand{\vmeanfluid}[1]{\left\langle {#1} \right\rangle_{\Yfluid}}

\newcommand{\gammaset}{{\left\{ a \cdot 10^b \ \unit{s^{-1}}| \ a=1,2,5; \ b = 1,2,3,4\right\} \cup \left\{10^5 \persec\right\}}}
\newcommand{\setD}{\Deff_{\gdot}}

\newcommand{\setAfluid}{\mathcal{\fluid{A}}}
\newcommand{\setNorth}{\mathcal{N}_{1}}
\newcommand{\setnoNorth}{\mathcal{N}_{2}}

\newcommand{\setFOT}{S_{\sf{T}}}
\newcommand{\setgdot}{S_{\gdot}}

\newcommand{\eoffmean}{e^{\sf{mean}}}
\newcommand{\eoffmeant}{e^{\sf{mean}}_{\sf{t}}}
\newcommand{\eoffmeanv}{e^{\sf{mean}}_{\sf{v}}}
\newcommand{\eon}{e_{\sf{on}}}
\newcommand{\eonmax}{e_{\sf{on}}^{\sf{max}}}
\newcommand{\eonmean}{e_{\sf{on}}^{\sf{mean}}}
\newcommand{\divrm}{{\mathrm{div}\ }}
\newcommand{\Psieff}{\bar{\Psi}}

\newcommand{\ffGamma}{{\mathpalette\makebbGamma\relax}}
\newcommand{\makebbGamma}[2]{%
  \raisebox{\depth}{\scalebox{1}[-1]{$\mathsurround=0pt#1\mathbb{L}$}}%
}
\newcommand{\symgrad}{\nabla^{\sf{s}}}

\newcommand{\sigmaop}{\mathcal{S}}
\newcommand{\twosphere}{\mathcal{S}^2}

\newcommand{\fzero}{\sty{0}}
\newcommand{\veczero}{\myvec{0}}
\newcommand{\myvec}[1]{\underline{#1}}
\newcommand{\mymat}[1]{\underline{\underline{#1}}}
\newcommand{\veca}{\myvec{a}}

\newcommand{\vecb}{\myvec{b}}

\newcommand{\vecd}{\myvec{d}}
\newcommand{\vecD}{\myvec{D}}
\newcommand{\vecDeff}{\myvec{\bar{D}}}

\newcommand{\vecn}{\myvec{n}}

\newcommand{\vecalpha}{\myvec{\alpha}}

\newcommand{\FOT}{\myvec{\lambda}}
\newcommand{\vecw}{\myvec{w}}

\newcommand{\vectau}{\myvec{\tau}}
\newcommand{\matA}{\mymat{A}}

\newcommand{\matH}{\mymat{H}}
\newcommand{\matN}{\mymat{N}}
\newcommand{\matW}{\mymat{W}}

\newcommand{\reals}{\mathds{R}}
\newcommand{\naturals}{\mathds{N}}
\newcommand{\setdata}{\mathcal{S}^{\sf{D}}}
\newcommand{\settrain}{\mathcal{S}^{\sf{t}}}
\newcommand{\setvalid}{\mathcal{S}^{\sf{v}}}

\newcommand{\charfuncfluid}{\fonems{\fluid{\chi_{\mysf{k}}}}}

\newcommand{\depthDMN}{\foneme{K}}
\newcommand{\clm}{\mathcal{C}_3}
\newcommand{\lmb}{\mathcal{B}^{\mysf{i}}_{\mysf{k}}}
\newcommand{\lm}{\mathcal{H}^{\mysf{i}}_{\mysf{k}}}
\newcommand{\Nweights}{N_{\sf{w}}}
\newcommand{\Nnormals}{N_{\sf{n}}}

\newcommand{\equalhat}{\mathrel{\stackon[1.5pt]{=}{\stretchto{%
    \scalerel*[\widthof{=}]{\wedge}{\rule{1ex}{3ex}}}{0.5ex}}}}

\newcommand{\Dmatrix}{\underline{\underline{\bar{D}}}}

\newcommand{\lossdmn}{\mathcal{L}}
\newcommand{\DMNLIN}{\mathcal{DMN}_{\sf{L}}}
\newcommand{\Nbatch}{N_{\sf{b}}}

\newcommand{\Nangles}{N_{\alpha}}
\newcommand{\otimessym}{\otimes_{\sf{s}}}
\newcommand{\spacejumps}{(\reals^3)^{\otimes \Nnormals}}
\newcommand{\spacefluct}{\left(\symdev\right)^{\Nweights}}
\newcommand{\psieffargs}{\left(\vecDeff + \matA \, \vecb\right)}

\newcommand{\Mco}{\mathcal{M}}

\newcommand{\mco}{m}

\newcommand{\fnr}{\fn_{\mysf{r}}}
\newcommand{\csfr}{c_{\mysf{r}}}
\newcommand{\ffMco}{\ffM}

\newcommand{\spacefourth}{(\reals^3)^{\otimes 4}}
\newcommand{\spacesecond}{(\reals^3)^{\otimes2}}
\newcommand{\jump}[1]{{\llbracket {#1} \rrbracket}}
\makeatletter
\def\@testdef #1#2#3{%
  \def\reserved@a{#3}\expandafter \ifx \csname #1@#2\endcsname
 \reserved@a  \else
\typeout{^^Jlabel #2 changed:^^J%
\meaning\reserved@a^^J%
\expandafter\meaning\csname #1@#2\endcsname^^J}%
\@tempswatrue \fi}
\makeatother
\makeatletter
\AtBeginDocument{%
    \immediate\write\@auxout{\catcode`^=12 }%
}
\makeatother
\title{\foneme{Deep material networks for fiber suspensions with infinite material contrast}}

\author[1]{Benedikt Sterr}
\author[1]{Sebastian Gajek}
\author[\fthrme{2}]{Andrew Hrymak}
\author[\fthrme{3}]{Matti Schneider}
\author[1]{Thomas Böhlke}

\affil[1]{Karlsruhe Institute of Technology (KIT), Institute of Engineering Mechanics}
\affil[\fthrme{2}]{University of Western Ontario, Department of Chemical and Biochemical Engineering}
\affil[\fthrme{3}]{University of Duisburg-Essen, Institute of Engineering Mathematics \newline correspondence to: \texttt{thomas.boehlke@kit.edu}}

\date{\today}
 
\begin{document}
\maketitle 
\begin{abstract}
\noindent
\ftwome{We extend the laminate based framework of direct Deep Material Networks (DMNs) 
to treat suspensions of rigid fibers in a non-Newtonian solvent. 
To do so, we derive two-phase homogenization blocks that are capable of treating \fthrms{incompressible}
fluid phases and infinite material contrast. In particular, we leverage
\fthrms{existing results for linear elastic laminates}
to identify closed form expressions for the linear homogenization functions of two-phase layered emulsions.
To treat infinite material contrast, we rely on the repeated layering of two-phase layered emulsions in the
form of coated layered materials. We derive \fthrms{necessary} and sufficient conditions \fthrms{which ensure that}
the effective properties of coated layered materials with \fthrms{incompressible} phases are non-singular, even if one of the 
phases is \fthrms{rigid}. 
\newline
\fthrms{With the }derived homogenization blocks and non-singularity conditions \fthrms{at hand},
we present a \fthrms{novel} DMN architecture, which we name the Flexible DMN (FDMN) architecture.
We build and train FDMNs to predict the effective stress response of shear-thinning 
fiber suspensions with a Cross-type matrix
material.
\ftwome{For 31 fiber orientation states, six load cases, and over a wide range of shear rates relevant to engineering processes,
the FDMNs achieve validation errors below 4.31\%
when compared to direct numerical simulations with Fast-Fourier-Transform based computational
techniques. \fthrms{Compared to a conventional machine learning approach introduced previously by the consortium of authors,}
FDMNs offer better accuracy at an increased computational cost for the considered material and 
flow scenarios.}

}
\end{abstract} 
 
\foneme{
\section{Introduction}
\label{sec:intro}
\subsection{State of the art} 

Especially because of their high specific stiffness, fiber 
reinforced composites are widely used for lightweight design, particularly in 
the transport and energy sectors~\cite{swolfs2017perspective,wazeer2023composites}. 
During the design process of fiber reinforced composite components,
combinations of molding simulations and structural simulations are central to facilitate use case appropriate
and material oriented design through digital twins~\cite{botin2022digital} and 
virtual process chains~\cite{henning2019fast,gorthofer2019virtual,meyer2023probabilistic}.
In molding simulations of fiber reinforce\ffouha{d} composite parts, 
accurate prediction of the local fiber suspension viscosity
is vital to correctly estimate process parameters~\cite{castro1990predicting,goodship2017arburg}
and other engineering quantities. In particular, the fiber orientation and the fiber volume distributions~\cite{tseng2018predictions},
the flow fields,~\cite{karl2021coupled}, as well as the final part properties~\cite{mortazavian2015effects,bohlke2019continuous,karl2021coupled}
depend on the suspension viscosity.
However, in component scale simulations, it is computationally infeasible to fully resolve the microstructure
of the whole domain of interest to compute the suspension viscosity.
For this purpose, analytical and computational homogenization methods are valuable tools to provide viscosity estimates
for molding simulations. 
However, significant challenges arise for the holistic analytical modelling of 
the suspension viscosity, because the suspension viscosity depends on the
local microstructure~\cite{karl2022unified}, the fiber volume 
fraction~\cite{sepehr2004rheological}, and the 
fiber orientation state~\cite{dinh1984rheological}. Additionally,
the flow field~\cite{cross1965rheology, sterr2023homogenizing} and the
melt temperature~\cite{williams1955temperature} influence the suspension viscosity as well.
Furthermore, typical matrix materials in fiber suspensions show non-Newtonian behavior, which 
adds additional complexity to the derivation of appropriate \ftwoms{models for the suspension viscosity}.

Depending on the fiber concentration of the suspension
and the behavior of the matrix material, different homogenization approaches have been proposed.
\ffouha{According to Pipes et al.~\cite[\S 1]{pipes1991constitutive} the fiber concentration regimes
may be defined using the fiber volume fraction~$\cf$ and the fiber aspect ratio~$\ra$
as follows. The dilute regime is defined through ${\cf<(1/\ra)^2}$,
the semi-dilute regime through ${(1/\ra)^2<\cf<1/\ra}$, the concentrated regime through ${1/\ra<\cf}$, and the hyperconcentrated regime through ${\ra>100}$
\cite{pipes1991constitutive}.}
Goddard~\cite{goddard1976tensile,goddard1976stress}, and
Souloumiac and Vincent~\cite{souloumiac1998steady} proposed self-consistent analytical models \ffouha{for different concentration regimes}.
The models by Goddard are restricted to the dilute and semi-dilute regimes, while
the models by Souloumiac and Vincent are applicable to the dilute, semi-dilute, and concentrated regimes.
Both models agree well with experimental results~\cite{mobuchon2005shear} qualitatively.
However, the quantitative prediction accuracy could be improved and varies strongly
depending on the shear rate and fiber volume fraction.
Similarly, a semi-analytical model proposed by F{\'e}rec et al.~\cite{ferec2016effect}
for Ellis and Carreau-type matrix behavior replicates the 
steady state flow solution of a simple shear flow, but its accuracy varies with
the shear rate and the fiber volume fraction.
By incorporating uniformly distributed fiber misalignments with
an orientation averaging approach~\cite{ericksen1960transversely,tucker1991flow},
Pipes et al.~\cite{pipes1991constitutive, pipes1992anisotropic, cofffin1991constitutive, pipes1994non}
derived models for the concentrated and hyper-concentrated regime. The 
models are only applicable to collimated fiber arrays, but \ftwoms{their} predictions
agree well with experimental studies by Binding~\cite{binding1991capillary}.
Combining orientation averaging with a deformation mode and microstructure
dependent informed isotropic viscosity, Favaloro et al.~\cite{favaloro2018new}
derived a model that aims to improve available molding simulation solvers
with only small modifications. Depending on the deformation mode and the
approximated anisotropic viscosity, the prediction quality of the model 
varies. However, molding simulations using the model successfully predicted
the shell-core effect that is common in a wide variety of fiber molding applications~\cite{tseng2018predictions,mortazavian2015effects}.
Overall, predicting the suspension viscosity accurately over the
wide variety of processing conditions used for engineering systems
proves difficult, which includes recent developments~\cite{tseng2021constitutive,khan2023constitutive}.

The development of homogenization methods is accompanied by considerable difficulties in studying
the fiber suspension viscosity experimentally. In rheometer studies, certain transient effects like fiber breakage and
fiber reorientation are difficult to quantify during the experiment. Thus, it is hard to link
a specific fiber orientation state and a specific fiber length distribution to a measured viscosity~\cite{binding1991capillary,schelleis2023optimizing}.
Furthermore, the measurement results may be affected by the interaction
between the fibers and the measurement devices~\cite{eschbach1993dynamic}.

In view of these difficulties, computational techniques offer \ftwoms{an extraordinary} approach
to \ftwoms{study the effective viscosity} of fiber suspensions
with non-Newtonian solvents by resolving the local fields of interest and providing insights, which are hard to obtain
otherwise. 
Combining a mass tracking algorithm for the free surface representation,
a lattice Boltzmann method for fluid flow, and an immersed boundary
procedure for the interaction between fluid and rigid particles,
{\v{S}}vec et al.~\cite{vsvec2012free} studied slump tests of rigid fibers and rigid spherical particles
suspended in a Bingham-type fluid. They compared 
slump tests of the suspension with a slump test of the pure matrix
material, and observed a smaller spread and an increased height in the test of the suspension material,
which implies an increased effective yield stress in the suspension.
Domurath et al.~\cite{domurath2020numerical} employed a Finite Element
Method (FEM) based approach to study the transversely isotropic fluid equation
by Ericksen~\cite{ericksen1960transversely}. They investigated the rheological
coefficients and found that the model by Souloumiac and Vincent~\cite{souloumiac1998steady}
overpredicts the orientation dependence of a rheological coefficient.
Extending work by Bert{\'o}ti et al.~\cite{bertoti2021computational} on suspensions with Newtonian solvents,
Sterr et al.~\cite{sterr2023homogenizing} used Fast Fourier Transform (FFT) based computational techniques~\cite{schneider2021review}
and the RVE method~\cite{kanit2003determination} to study the effective viscosity of fiber suspensions with non-Newtonian solvents.
They investigated the effects of anisotropic shear-thinning on the effective suspension viscosity 
for varying fiber volume fractions, shear rates, and flow scenarios.

Accurate computational predictions of the microscopic behavior of a composite often require significant computational 
effort~\cite{renard1987etude,feyel2003multilevel,spahn2014multiscale,kochmann2016two,kochmann2018efficient}, 
which motivates the use of fast data based surrogate models for component scale simulations~\cite{gajek2021fe}.
Ashwin et al.~\cite{ashwin2022deep,ashwin2024physics} trained a Multi-Layer-Perceptron,
a Convolutional Neural Network, and a U-Net~\cite{ronneberger2015u} on particle 
resolved simulation data to predict fluid forces in dense ellipsoidal
particle suspensions. They restricted to Newtonian matrix behavior and trained the
networks on data for various Reynolds numbers and fiber volume fractions.
Depending on the considered combination of Reynolds number and particle
volume fraction, the prediction accuracy of the surrogate models varies.
Boodaghidizaji et al.~\cite{boodaghidizaji2022multi} use 
a multi-fidelity approach with neural networks and Gaussian processes 
to predict the steady state viscosity of fiber suspensions with a Newtonian 
solvent. To form the training data set, they combine low-fidelity estimates from constitutive equations
with high-fidelity data obtained from numerically solving the involved partial differential equation system. 
The prediction accuracy of the multi-fidelity neural network and the multi-fidelity Gaussian process
for simple shear flow depends strongly on the investigated parameters, especially the fiber volume fraction.
Sterr et al.~\cite{sterr2024machine} derived 
four models for fiber suspensions with a Cross-type matrix fluid
by combining FFT-based \ftwoms{computational} homogenization techniques 
with supervised machine learning.
\ftwoms{They} investigated the anisotropic shear-thinning characteristics
of the suspension viscosity for a variety of fiber orientation states \ftwoms{via computational experiments},
and formulated model candidates based on the observed phenomena.
Using supervised machine learning techniques, they identified the model parameters from
computational data, so that three of the four models were able to predict the 
fiber suspension viscosities to engineering accuracy for a wide range of engineering load cases.
Generally, in addition to accurate estimation, extrapolation beyond the training data, as well as ensuring 
thermodynamical consistency prove challenging in the construction of surrogate models.

In the context of solid materials without kinematic constraints,
Liu et al.\cite{liu2019deep, liu2019exploring} proposed Deep Material Networks (DMNs) as 
surrogate models for the \fthrms{full-field computational homogenization of microstructured materials}.
\ftwoms{Their approach is based on nesting rotated laminates in an $N$-ary tree structure of $N$-phase laminates,
and thus constructing a micromechanically motivated deep learning architecture.
The \fthrms{volume fractions} and rotations of the DMN are then identified via supervised machine learning
on linear elastic data. Remarkably, even if a DMN is trained on data obtained by solving linear homogenization problems, the
predictions of the DMN for non-linear homogenization problems are impressively accurate. 
Gajek et al.~\cite{gajek2020micromechanics}
further developed the DMN architecture into the rotation free direct DMN architecture.
Direct DMNs feature a faster and more robust training process, as well as
an efficient evaluation scheme for non-linear problems.
Gajek et al.~\cite{gajek2020micromechanics} also showed that
the thermodynamic consistency of the laminates is preserved in (direct) DMNs, such that the
resulting DMN is thermodynamically consistent as well. 
Furthermore, Gajek et al.~\cite{gajek2020micromechanics} proved that non-linear homogenization is determined by
linear homogenization to first order for two-phase materials.
In a later article, Gajek et al.~\cite{gajek2021fe} used direct DMNs to accelerate two-scale
FE simulations of fiber reinforced composites by 
augmenting direct DMNs with the fiber orientation interpolation concept introduced by K\"obler et al.~\cite{kobler2018fiber}.
\fthrms{Alternatively, DMN parameters may be interpolated with regard to microstructural parameters by 
a posteriori interpolation as proposed by 
Liu et al.~\cite{liu2019transfer} and Huang et al.~\cite{huang2022microstructure},
or by augmenting DMNs with
neural networks as proposed by Li~\cite{li2024micromechanics}.}
To leverage the capabilities of direct DMNs in concurrent thermomechanical two-scale simulations of composite components\fthrme{,}
Gajek et al.~\cite{gajek2022fe} further extended the direct DMN architecture to incorporate thermomechanical coupling. 
\fthrms{Also in a thermomechanical setting, Shin et al.~\cite{shin2024deep} trained DMNs on linear thermoelastic data instead
of linear elastic data, which improved the quality of fit for the effective thermal expansion properties,
but only slightly affected the non-linear prediction quality of the DMNs.}
\fthrms{Additionally, they employed DMNs for uncertainty quantification, and for the inverse problem of optimizing a thermal boundary
condition to achieve a desired thermo-elasto-viscoplastic response.}
\ftwome{By developing an inelastically-informed training strategy for DMNs, Dey et al.~\cite{dey2022training}
successfully predicted the creep behavior of fiber reinforced thermoplastics, which involves multiple scales in both space and time.
This enabled the inverse calibration of parameters for creep and plasticity \fthrtb{constitutive equation}s by using DMNs as surrogates
for otherwise costly FFT-based computations~\cite{dey2023rapid}. Furthermore, Dey et al.~\cite{dey2024effectiveness}
leveraged DMNs in combination with fiber orientation interpolation~\cite{kobler2018fiber}
to characterize the behavior of fiber reinforced thermoplastics including damage, plasticity, and creep.}
\fthrms{Overall, DMNs were extended and applied to treat a wide variety of problems, such as
interface damage~\cite{liu2020deep}, the modeling of multiscale strain
localization~\cite{liu2021cell}, problems involving woven materials~\cite{wu2021micro} and
porous materials~\cite{noels2022interaction}, as well as the architecture independent
treatment of multi-phase composites~\cite{noels2022micromechanics}.}

For composites with kinematically unconstrained solid phases,
the presented literature provides compelling evidence for the accuracy and versatility of (direct) DMNs.
However, the (direct) DMN architecture has yet to be applied to composites involving fluids or infinite material contrast,
which is required for suspensions of rigid fibers. Because (direct) DMNs are based on laminates of solids,
the treatment of fluids with a DMN architecture requires a different type of building block.
Additionally, treatment of infinite material contrast with DMNs is challenging,
because the rank-one laminates of a DMN have singular effective properties
if one of the phases is singular, i.e., \fthrms{rigid}. Therefore,
the singular effective properties of the rank-one laminates may propagate through
the whole DMN during the training \fthrme{and} evaluation processes, rendering the effective properties of the DMN singular as well.  
In this article we address these issues and propose an architecture to treat suspensions of rigid fibers.
}
}

\subsection{Contributions}
\foneme{
\ftwoms{
\ftwoms{
}In this work, we extend the direct DMN architecture for kinematically unconstrained solid phases with finite material contrast 
to an architecture that is able to treat 
\ftwoms{suspensions of rigid fibers.
In particular, this \fthrms{task} requires the treatment of \fthrms{fluid} phases \fthrme{and} infinite material contrast.
We name this architecture the Flexible DMN (FDMN) architecture.
To treat} fluid materials with FDMNs, we derive homogenization blocks
for layered emulsions that are governed by Stokes flow,
and consist of possibly incompressible phases with finite material contrast.
In section~\ref{subsec:affinefields}, we \fthrme{establish} that the velocity field
in this type of layered material is phase-wise affine,
if the dissipation potentials of the phases are strictly rank-one convex.
We use this result and follow an analytical procedure \fthrms{described by Milton~\cite[§9]{milton2022theory}} to derive
the closed form linear homogenization function of the considered layered
emulsions in section~\ref{subsec:veff_milton}. To treat infinite material contrast, we study
a particular type of layered material, namely, coated layered materials (CLMs), in section~\ref{subsec:coated_layered_materials}.
We obtain closed form homogenization functions for CLMs with kinematically unconstrained or incompressible phases.
With the objective of using CLMs as FDMN building blocks in the presence of \ftwome{rigid fibers}, 
we derive conditions for the required number of layering directions and their orientation,
such that the effective properties of CLM are non-singular. 
We do so for \fthrms{CLMs with incompressible and rigid phases}.}

\ftwome{In section~\ref{sec:identifying_dmns}, we present the FDMN architecture
as an extension of the (direct) DMN architecture. An FDMN arises by replacing
the rank-one laminates of a (direct) DMN with rank-one layered materials capable of treating fluids,
and non-singular CLMs capable of treating infinite material contrast. Also, we discuss the
material sampling, the offline training procedure, as well as the online evaluation \ftwome{of FDMNs}
in the context of incompressible rigid fiber suspensions.}
\foneme{\ftwome{We apply} FDMNs to predict the viscous response 
of fiber polymer suspensions \ftwome{in section~\ref{sec:rigid_fibers_in_pa6}}, demonstrating both the ability to handle incompressible fluid phases and infinite material contrast.
We use FFT-based \ftwoms{computational} homogenization techniques to generate linear training data and non-linear
validation data for shear-thinning fiber suspensions with a Cross-type matrix behavior, and a fiber volume fraction of 25\%.
We consider 31 different fiber orientation states and a variety of incompressible elongational and shear
flows. Using the computational data,
we train FDMNs for all fiber orientation states, and study the prediction accuracy of the FDMNs in the non-linear case.
For all investigated load cases and microstructures, the FDMNs achieve validation errors below 4.31\%
over a wide range of shear rates relevant to engineering processes. \ftwome{\fthrms{Not considering the time required
to generate the training data and to train an FDMN}, the
FDMNs achieve speedup factors between 11785 and 17225 compared to FFT-based computations. Finally, we compare
the prediction accuracy of the FDMN based approach with a different machine learning approach by Sterr et al.~\cite{sterr2024machine},
and find that FDMNs offer better accuracy and flexibility at a higher computational cost for the considered material and flow
scenarios.} 
}
}

\section{Homogenization of layered emulsions with infinite material contrast}\label{sec:homogenization_of_emulsions}

\subsection{Phase-wise affine displacement and velocity fields in laminates and emulsions}\label{subsec:affinefields}
\ftwoms{
Like for (direct) DMNs, the building blocks of an FDMN 
should \fonems{admit linear homogenization functions in closed form}, and there
should exist efficient solution schemes to compute the stress response
of the building blocks \fonems{in case} their phases are non-linear.
In analogy to the laminate based architecture of the (direct) DMN, we 
thus consider layered emulsions as the building blocks of the FDMN architecture.
Like laminates, layered emulsions consist of multiple,
possibly non-linear, fluid materials, that are arranged 
such that every phase boundary between the phases is 
orthogonal to the layering direction~\foneme{$\fn\in\twosphere$, on the 2-sphere~$\twosphere$}.}
For the offline training and \fonems{the} online evaluation of an FDMN with layered emulsions as building blocks, 
\fonems{knowing} the homogenization functions of layered emulsions with linear and non-linear phases is necessary. Primarily,
we look for similarities \fonems{between} the homogenization of layered emulsions and \fonems{solid} laminates
that can be leveraged.  
\fonems{In laminates, the displacement fields \fonems{are} phase-wise affine,
and we wish to know if the velocity fields in layered emulsions belong to a 
particular class of fields as well.} 
If the velocity fields \fonems{are} phase-wise affine, established solution
techniques on (direct) DMN architectures~\cite{gajek2020micromechanics} can be employed for the online evaluation
of the FDMN, and convenient linear homogenization equations can be employed for the offline training.
\fthrms{We follow a procedure detailed by Kabel et al.~\cite{kabel2016model} on the homogenization
of laminates, and establish the existence and uniqueness of phase-wise affine velocity fields in a particular type of emulsion.}
\fonems{We}
consider a $K$-phase layered emulsion with layering direction~$\fn$, which, similar to a laminate,
is constructed by arranging $K$-phases so that the direction~$\fn$ is orthogonal to all phase boundaries.
The emulsion occupies a \fthrms{rectangular and periodic} volume~$\Yfluid\subset \reals^3$, and consists of \fonems{non-Newtonian} phases with
rank-one convex dissipation potential densities ${\dpotk:\spacesecond\rightarrow\reals}$\ftwoms{, such that
\begin{equation}
    \dpotk(\fL + \beta\fa \otimes \fb) \leq \beta \dpotk(\fL + \fa \otimes \fb) + (1 - \beta)\dpotk(\fL)
    \quad \forall \, \beta \in [0, 1], \, \fa, \fb \in \reals^3\backslash \{\fzero\},
\end{equation}
where~$\fL\in\spacesecond$ denotes the velocity \fthrtb{gradient}\ftwotb{, and the operator~$(\cdot)^{\otimes a}$ 
constructs a tensor space of $a$-th order.}
Also},~$\Yfluid_{\mysf{k}}\subset \Yfluid$ denotes the sub volume of the $\mysf{k}$-th \fonems{phase}\fthrme{,}
and the \fonems{phases} do not intersect, i.e.,
\begin{equation}
    \Yfluid_{\mysf{\fthrms{j}}} \cap \Yfluid_{\mysf{\fthrms{k}}} = \emptyset \quad\text{if} \quad \fthrms{\mysf{j} \neq \mysf{k}},
\end{equation}
\fthrms{and form the volume~$Y$, such that
\begin{equation}
    Y = \bigcup_{j=1}^K Y_j.
\end{equation}
}
\ftwoms{Let the operator~$\vmean{\cdot}$ denote volume averaging \fthrms{of a quantity}
over a volume~$Y$, \fthrms{i.e.,}
\begin{equation}
    \vmean{\cdot} \equiv \frac{1}{|Y|}\int_Y (\cdot) \dif \fthrtb{Y(\fx)} \quad \text{with} \quad \ |Y| \equiv \int_Y \dif \fthrtb{Y(\fx)}.
\end{equation}
Then,} we express the velocity field~$\fv:\Yfluid\rightarrow\reals^3$ as
\begin{equation}\label{eq:velosplit}
    \fv = \foneme{\Leff}\fx + \fvhat, \quad \text{with} \quad \foneme{\vmeanfluid{\fthrtb{\nabla\fvhat}}} = \fzero,
\end{equation}
where~\foneme{$\Leff\in\spacesecond$ stands for the prescribed effective velocity gradient} \fthrtb{and}
~$\fvhat$ is the velocity fluctuation field\fthrtb{.}
Accordingly, the \fthrms{periodic} dissipation potential density function \foneme{${\Psi:\Yfluid\times\spacesecond\rightarrow\reals}$} of the emulsion is defined through
\begin{equation}\label{eq:defpsi}
    \Psi(\fx,\symgrad\fv) = \sum_{i}^{K} \charfuncfluid(\fx)\dpotk(\fx, \symgrad\fv),
\end{equation}
where~$\charfuncfluid:\foneme{\Yfluid}\rightarrow\{0,1\}$ \fthrms{refers to} the characteristic 
function of the \fonems{sub volume~$\Yfluid_{\mysf{k}}$}\fthrtb{, and~$\symgrad$
stands for the symmetrized gradient}.
\fonems{Let the flow inside the emulsion be governed by Stokes flow, i.e., 
the advectorial forces in the layered emulsion are small in contrast to
the viscous forces, then the \ftwoms{steady state balance of mass
\begin{equation}\label{eq:massbalance_co}
    \divrm{(\rho\fv)} = 0,
\end{equation}
where~$\rho:\Yfluid:\rightarrow\reals$ denotes the mass density field, is satisfied.}
Additionally, the stress field
\begin{equation}
    \fsigma = \sigmaop(\cdot, \Deff + \nabla^{\mysf{s}} \fvhat)
\end{equation}
in terms of the effective strain rate tensor~\fthrms{$\Deff=(\Leff+\T{\Leff})/2$}, and the associated stress 
operator\fthrme{~$\sigmaop$}
\ftwoms{on the space of symmetric second order tensors~$\sym\subset\fthrms{\spacesecond}$}\fthrtb{~\cite{edelen1973existence}}
\begin{align}
    \sigmaop: \Yfluid \times \sym &\rightarrow \sym,\\
    (\fx, \fD) &\mapsto {\pdv{\Psi}{\fD}} {(\fx,\fD)},
\end{align}
satisfies the balance of linear momentum without inertial effects \fonetb{and volume forces}
\begin{equation}\label{eq:momentumbalance}
    \ftwoms{\divrm{\fsigma} = \fzero.}
\end{equation}
Therefore, the effective response of the emulsion is obtained from the minimization problem
\begin{equation}\label{eq:minemulsion}
    \vmeanfluid{\Psi(\fx, \Deff+\symgrad\fvhat)} \longrightarrow \min_{\hat{\bm{v}} \in \setAfluid},
\end{equation}
where~$\setAfluid$ stands for the admissible set
\begin{equation}\label{eq:setAfluid}
    \setAfluid = \left\{\fvhat:\Yfluid\rightarrow\reals^3 \quad \middle|\quad  \ftwoms{\divrm{(\rho\fv)}} = 0 \quad \text{and} \quad \vmean{\fthrtb{\nabla\fvhat}} = \fzero\right\}
\end{equation}
Hence, assuming there is no slip between the phases, 
the proof by Kabel et al.~\cite[§2]{kabel2016model} extends to 
layered emulsions \ftwoms{with finite material contrast}. \ftwoms{Therefore}, if the dissipation potential densities~$\Psi_{\mysf{k}}$ are strictly rank-one convex, there
exists a unique minimizer~$\fvhat$ in the class of phase-wise affine velocity fields. 
\ftwoms{As a special case, this is also true for incompressible Stokes flow\fthrtb{s}, for which the balance of mass \fthrtb{simplifies to}
\begin{equation}\label{eq:massbalance_ico}
    \fthrms{\divrm{\fv}} = 0.
\end{equation}
}%
}In the following, we use the term \textit{layered emulsion} 
only for the type of emulsion defined above.
\subsection{\ftwome{Two-phase layered emulsions with Newtonian phases}}\label{subsec:veff_milton}

Similar to two-phase laminates of linear elastic solid materials, we wish to find a
convenient closed form expression for the linear homogenization function 
of two-phase layered emulsions.
With the goal of using layered emulsions
as building blocks of an FDMN, we dedicate this section
to deriving such a closed form expression for two-phase layered emulsions with \ffouha{kinematically unconstrained or} incompressible Newtonian phases.
For this purpose, we employ the results of the previous section~\ref{subsec:affinefields},
and extend an existing homogenization approach for two-phase laminates of linear elastic phases
by Milton~\cite[§9]{milton2022theory}. 
\ftwoms{To emphasize
the link between the linear homogenization of laminates
and layered emulsions we define the class of linear materials~$\Mco$,
and study layered materials of solids and fluids in parallel.
A material~$m_{\mysf{k}}$ in the class of linear materials~$\Mco$ is characterized by the \fonems{primal}
material \ftwoms{tensor~$\ffMco_{\mysf{k}}$} \fonems{or its dual tensor~\ftwoms{$\ffK_{\mysf{k}}\equiv \ffMco_{\mysf{k}}^{-1}$}}, such that
\begin{equation}
    (\ffMco_{\mysf{k}},\ffK_{\mysf{k}}) \in \{(\symplusfour)^2, (\symdevplusfour)^2\}.
\end{equation}
Here, the convex cones~$\symplusfour\subset\spacefourth$ and~$\symdevplusfour\subset\spacefourth$ comprise all 
fourth order tensors~$\fX\in\spacefourth$, which are positive definite on the vector 
spaces of symmetric second order tensors~$\sym$ and traceless symmetric second order tensors~$\symdev\subset\sym$, respectively.
Additionally, a tensor~$\ffX\in\{\symplusfour,\symdevplusfour\}$ has minor and major symmetries
\begin{equation}\label{eq:minmajsyms}
    \ffX = \ffX^{\fthrtb{\sf{T_H}}}=\ffX^{\fthrtb{\sf{T_L}}}=\ffX^{\fthrtb{\sf{T_R}}},
\end{equation}
and a tensor~$\ffX\in\symdevplusfour$ maps the second order unit tensor~$\fI$ to zero, such that
\begin{equation}
    \ker\ffX = \fI, \quad\text{i.e.,}\quad \ffX[\fI] = \fzero \fthrtb.
\end{equation}
\fthrtb{Here, we define the map~$\ffT[\fY]$ in component form and in the standard basis of~$\reals^3$ as
\begin{equation}
    \ffT[\fY] \equalhat T_{ijkl}Y_{kl}, \quad \text{for some}\quad \fY\in\spacesecond,\ffT\in\spacefourth.
\end{equation}
}Any
material~$\mco_{\mysf{k}} \in \Mco$ follows the \fthrtb{linear constitutive equations}
\begin{equation}\label{eq:sigmaco}
    \fsigma^{\mysf{co}} = \ffMco_{\mysf{k}}[\symgrad\fw\fthrms{]} \fonems{\quad \text{and} \quad \symgrad\fw = \ffK_{\mysf{k}}[\fsigma^{\mysf{co}}]},
\end{equation}
if the material~$m_{\mysf{k}}$ is \fonetb{not subjected to kinematic constraints}, or the \fthrtb{constitutive equation}
\begin{equation}\label{eq:sigmaico}
    \fsigma^{\mysf{ico}} \fonems{= \ftau - p\fI} = \ffMco_{\mysf{k}}[\symgrad\fw] - p \fI, \fonems{\quad \text{and} \quad \symgrad\fw = \ffK_{\mysf{k}}[\ftau]},
\end{equation}
if the material~$m_{\mysf{k}}$ is incompressible.
Here, $\fw$ denotes the field associated with the type of material~$m_{\mysf{k}}$, i.e., the 
displacement field~$\fu$ for linear elastic solids or the velocity
field~$\fv$ for linearly viscous fluids. For the \fthrtb{constitutive equation}~\eqref{eq:sigmaico}
we use an additive split of the stress field~$\fsigma$ 
\begin{equation}\label{eq:stresssplit}
    \fsigma = \ftau - p\fI,
\end{equation}
into the \fthrtb{deviatoric} stress field~$\ftau:\Yfluid\rightarrow\symdev$ and the pressure field~$p:\Yfluid\rightarrow\reals$.
Here, we use~$Y\subset\reals^3$ to denote the volume occupied by a two-phase layered material, i.e., a laminate or a layered emulsion,
and~$Y_1,Y_2\subset Y$ to denote the volume occupied by the two phases~$m_1,m_2\in\Mco$ of the layered material.
We define the effective properties~$(\Meff,\Keff)\in\{(\symplusfour)^2,(\symdevplusfour)^2\}$ of the layered material 
in the sense of the Hill-Mandel condition~\cite{hill1963elastic,mandel1966contribution}
\begin{equation}~\label{eq:hillmandel}
    \vmean{\fsigma\cdot\symgrad\fw} = \vmean{\fsigma} \cdot \vmean{\symgrad\fw} \equiv \Meff\cdot\left(\vmean{\symgrad\fw} \otimes \vmean{\symgrad\fw}\right)\fthrtb{.}
\end{equation}
Then, for perfectly bonded laminates of linear solid materials without kinematic constraints,
a closed form expression for the primal or dual effective properties of the laminate~$\Aeff\in\{\Meff,\Keff\}$ can be derived
using the procedure detailed by Milton~\cite[§9]{milton2022theory}, such that
\begin{equation}\label{eq:Aeff_milton}
    \left(\left(\Aeff - \ffA_0\right)^{-1} + \ftwoms{\GammaAzero}\foneme{(\fn)}\right)^{-1} = \left\langle\left(\left(\ffA(\fx) - \ffA_0\right)^{-1} + \ftwoms{\GammaAzero}\foneme{(\fn)}\right)^{-1}\right\rangle.
\end{equation}
Here,\fthrms{ the inverse is taken on the space~$\symplusfour$ in which~$\Aeff\in\symplusfour$ lives,}~$\ffA:Y\rightarrow\symplusfour$ denotes the field of the phase wise primal or dual material properties
\begin{equation}
    \ffA: Y \rightarrow \foneme{\symplusfour}, \quad \fx \mapsto 
        \begin{cases}
            \ffA_1 & \text{if $\fx \in Y_1$}\fthrms{,} \\
            \ffA_2 & \text{if $\fx \in Y_2$}\fthrms{,}
        \end{cases}
\end{equation} 
\fthrms{and}~$\ffA_0\in\symplusfour$ refers \fthrme{to} an arbitrary reference material tensor \fthrms{such that the
term~$(\ffA(\fx) - \ffA_0)$ is invertible. Also, the} the operator~$\fthrms{\GammaAzero}(\fn)\in\spacefourth$ is defined as
\begin{equation}\label{eq:Gammadef}
    \ftwoms{\GammaAzero}(\fn) = \GammaA(\fn)\left[\GammaA(\fn)\ffA_0\GammaA(\fn)\right]^{\fthrms{\dagger}}\GammaA(\fn),
\end{equation}
where\fthrms{~$(\cdot)^{\dagger}$ stands for the Moore--Penrose pseudoinverse}\fthrme{,
and the operators~$\GammaAzero(\fn)$ and~$\GammaA(\fn)$ project onto the same subspace with respect
to different inner products. More specifically, the operator~$\GammaAzero(\fn)$ is an orthogonal projector with
respect to the~$\ffA_0$-weighted inner product
\begin{equation}
    \langle\fX,\fX\rangle_{\ffA_0} = \fX\cdot\ffA_0[\fX], \quad \fX \in\sym,
\end{equation}
and the operator~$\GammaA(\fn)$ is an orthogonal projector with respect to the Frobenius inner product
\begin{equation}
    \langle\fX,\fX\rangle_{\sf{F}} = \fX\cdot\ffI[\fX] = \fX\cdot\fX, \quad \fX \in\sym,
\end{equation}
where~$\ffI\in\spacefourth$ denotes the fourth order identity tensor.
}The
operator~$\GammaA(\fn)$ encodes restrictions imposed by the
periodicity of the laminate, the perfect bonding of the phases, the momentum balance~\eqref{eq:momentumbalance},
and the \fthrtb{constitutive equation}~\eqref{eq:sigmaco}.
Depending on whether we wish to compute the effective primal or the effective dual properties
of the laminate, a different operator~$\GammaA(\fn)$ is required.
For the homogenization of the primal effective properties, i.e.,~$\Aeff=\Meff$, the operator~$\GammaA(\fn)$
equals the operator~$\Gammasolidone(\fn)$.
For the homogenization of the dual effective properties, i.e.,~$\Aeff=\Keff$, the operator~$\GammaA(\fn)$
equals the operator~$\Gammasolidtwo(\fn)$.
The operators~$\Gammasolidone$ and~$\Gammasolidtwo$, are defined through their action on a second order tensor~$\fX\in\sym$
\begin{equation}
    \Gammasolidone(\fn)[\fX] = \foneme{2(\fX\fn)\otimessym}\fn - (\fn\cdot\fX\fn)\fn\otimes\fn 
    \quad \text{\fthrms{and}} \quad
    \Gammasolidtwo(\fn)[\fX] = \fX - \Gammasolidone(\fn)\fX,
\end{equation}
\fthrtb{where~$\otimessym$ denotes the symmetrized dyadic product.}{ The operators~$\Gammasolidone$ and~$\Gammasolidtwo$ 
project onto the subspaces}\begin{equation}
    \spacestrainco = \left\{\foneme{\sym} \ni \fX = \fd \otimes \fn + \fn \otimes \fd \quad\middle|\quad \fd \in \reals^3\right\} 
    \quad \text{and} \quad
    \spacestressco = \left\{\foneme{\sym} \ni \fX \quad\middle|\quad \fX\fn = \fzero \right\}\fthrtb{,}
\end{equation}
\fthrtb{and} 
arise from the fact that the jumps of the 
field~$\jump{\symgrad\fw}\in\spacestrainco$ 
\begin{equation}\label{eq:jump_w_co}
    \jump{\symgrad\fw} = \jump{\nabla\fw\,\fn}\otimessym\fn,
\end{equation}
and the jump of the stress~$\jump{\fsigma}\in\spacestressco$ live on mutually orthogonal
subspaces~\cite[§9]{milton2022theory}, because it holds that \nolinebreak
\begin{equation}\label{eq:jump_t_co}
    \jump{\fsigma\fn} = \fzero.
\end{equation}
\fthrtb{Here, we define the jump~$\jump{q}$ of a quantity~$q$ across a phase boundary with normal vector~$\fn$ as
\begin{equation}
    \jump{q} = q^{+} - q^{-},
\end{equation}
where~$q^{+}$ and~$q^{-}$ denote the outer and inner limits of the quantity~$q$ with regard to the outside
facing normal vector~$\fn$, respectively. 
}

Interestingly, the constitutive equations for kinematically unconstrained linear elastic solids and linearly viscous fluids
share the same structure, see equation~\eqref{eq:sigmaco}. Additionally, equations~\eqref{eq:jump_w_co}
and~\eqref{eq:jump_t_co} are satisfied for layered emulsions by definition, because there is no
slip between the phases, and the momentum balance~\eqref{eq:momentumbalance} is satisfied.
Thus, the homogenization problem for the considered layered emulsions has the same structure as the homogenization problem for laminates.
As a result, the effective properties~$\Aeff\in\symplusfour$ of a two-phase layered emulsion of kinematically unconstrained materials~$m_1,m_2\in\Mco$
can be computed with equation~\fthrms{\eqref{eq:Aeff_milton}}.
If the kinematic field~$\fw$ is solenoidal\fthrms{ s}uch that the incompressible balance of mass~\eqref{eq:massbalance_ico}
is satisfied, it follows from equation~\eqref{eq:jump_w_co} that
\begin{equation}\label{eq:jump_ico}
    \jump{\symgrad\fw\,\fn}\cdot\fn = 0.
\end{equation}
Also, if the materials~$m_1,m_2\in\Mco$ are incompressible, it follows from equations~\eqref{eq:stresssplit} and~\eqref{eq:jump_t_co} that
\begin{equation}
    \jump{\fsigma\fn} = \fthrms{\fzero} \quad \iff \quad \jump{\ftau\fn} = \jump{p}\fn.
\end{equation}
Therefore, the jumps~$\jump{\symgrad\fw}\in\spacestrainico$ and~$\jump{\fsigma}\in\spacestressico$ live in the mutually orthogonal
spaces
\begin{equation}\label{eq:spacestrainico}
    \spacestrainico = \left\{\foneme{\symdev} \ni \fX = \foneme{2\,\fd \otimessym \fn} - 2(\fd\cdot\fn)\fn\otimes\fn
    \quad\middle|\quad \fd \in \reals^3\right\} 
\end{equation} 
and
\begin{equation}\label{eq:spacestressico}
    \spacestressico = \left\{\foneme{\symdev} \ni \fX \quad\middle|\quad \fX\fn = a\fn, \quad a\in\reals \right\}.
\end{equation}
The two operators~$\Gammafluidone(\fn)$ and~$\Gammafluidtwo(\fn)$, defined through their action on a traceless second order tenso\fthrme{r}
${\fX\in\ftwoms{\symdev}}$\fthrms{,}
\begin{equation}\label{eq:gammasfluid}
    \Gammafluidone(\fn)[\fX] = \foneme{2(\fX\fn)\otimessym}\fn - 2(\fn\cdot\fX\fn)\fn\otimes\fn \quad \text{\fthrms{and}} 
    \quad \Gammafluidtwo(\fn)[\fX] = \fX - \Gammafluidtwo(\fn)[\fX],
\end{equation}
project onto the spaces~$\spacestrainico$ and~$\spacestressico$, respectively. 
Consequently, the procedure detailed in Milton~\cite{milton2022theory}
can be used to compute the effective properties~$\Aeff\in\symdevplusfour$ of a layered material
with incompressible constituents. In this context the homogenization equation~\eqref{eq:Aeff_milton} applies if
the property field~$\ffA:Y:\rightarrow\symdevplusfour$, the reference properties~$\ffA_0\in\symdevplusfour$,
and the operator~$\GammaA(\fn)\in(\Gammafluidone(\fn),\Gammafluidtwo(\fn))$ \fthrms{\fthrme{are} selected based on
whether the tensor~$\Aeff$ represents primal or dual effective properties}.
In summary, for a layered material consisting of linear materials~$m_1,m_2\in\Mco$,
the effective material properties~$\ftwoms{(\Meff,\Keff) \in \{(\symplusfour)^2, (\symdevplusfour)^2\}}$ can be 
computed with the homogenization equation~\eqref{eq:Aeff_milton}.
The choice of the operator~$\GammaA(\fn)$ depends on whether the phases are compressible or incompressible,
and whether the primal or dual effective tensors are sought. 
}

\fonems{
\subsection{Coated layered materials}\label{subsec:coated_layered_materials}
}\fonems{
\ftwoms{
Because we are interested in building FDMNs for suspensions of rigid particles,
the architecture of the FDMN should be capable of handling infinitely viscous, i.e., rigid,
phases. In this section, we leverage the homogenization equation~\eqref{eq:Aeff_milton} to study the
homogenization of repeatedly layered materials in the presence of \fthrms{rigid} phases. }\ftwoms{We
define and use the term \textit{rank} of a layered material in the following
to better distinguish between the possible ways to construct layered materials. 
The rank~$R\in\naturals_{\geq0}$ of a \fonems{hierarchical}, layered
material defines the number of layering steps that are needed to construct the
material using layering directions~\foneme{$\fnr\in\twosphere$}, such that the index~${\mysf{r}}$
starts at one and ends at~$R$, i.e.,~${\mysf{r}}\in\{1,...,R\}$. By definition,
a rank-$0$ layered material is simply a single phase material without any layering.
Also, the materials that are layered at each layering step need not be the same.
For example, let the first layer in a rank-2 layered material consist of the materials~$m_1$
and~$m_2$. Then, the second layer may consist of the materials~$m_1$ and~$m_2$
in the form of the first layer, and a third material~$m_3$.}

Rank-one two-phase layered materials
are unsuitable as DMN building blocks in the case of infinite material contrast, because
their effective properties are always singular if one phase is singular.
In search of a building block capable of handling \fthrms{rigid phases},
we study the effective properties of a particular type of layered material following 
the terminology and \ftwome{the} \fonems{analytical} approach \fthrme{detailed} by Milton~\cite[§9]{milton2022theory}.
}
We refer to this particular type of layered materials as coated layered materials (CLMs). To construct
a CLM, one material~$m_2$, called the coating, is repeatedly layered onto another material~$m_1$, called the core.
The layering process is divided into multiple steps, such that the $\mysf{r}$-th layering step is applied in direction~$\fnr$,
and the total number of layering steps equals the rank~$R$ of the CLM.
\begin{figure}
    \centering
    \begin{subfigure}[t]{0.49\textwidth}
        \centering
        \vskip 0pt
        \begin{tikzpicture}[spy using outlines={circle,yellow,magnification=5,size=1.5cm, connect spies}]
            \node[anchor=south west] (image) at (0,0) {\pgfimage[width=\textwidth]{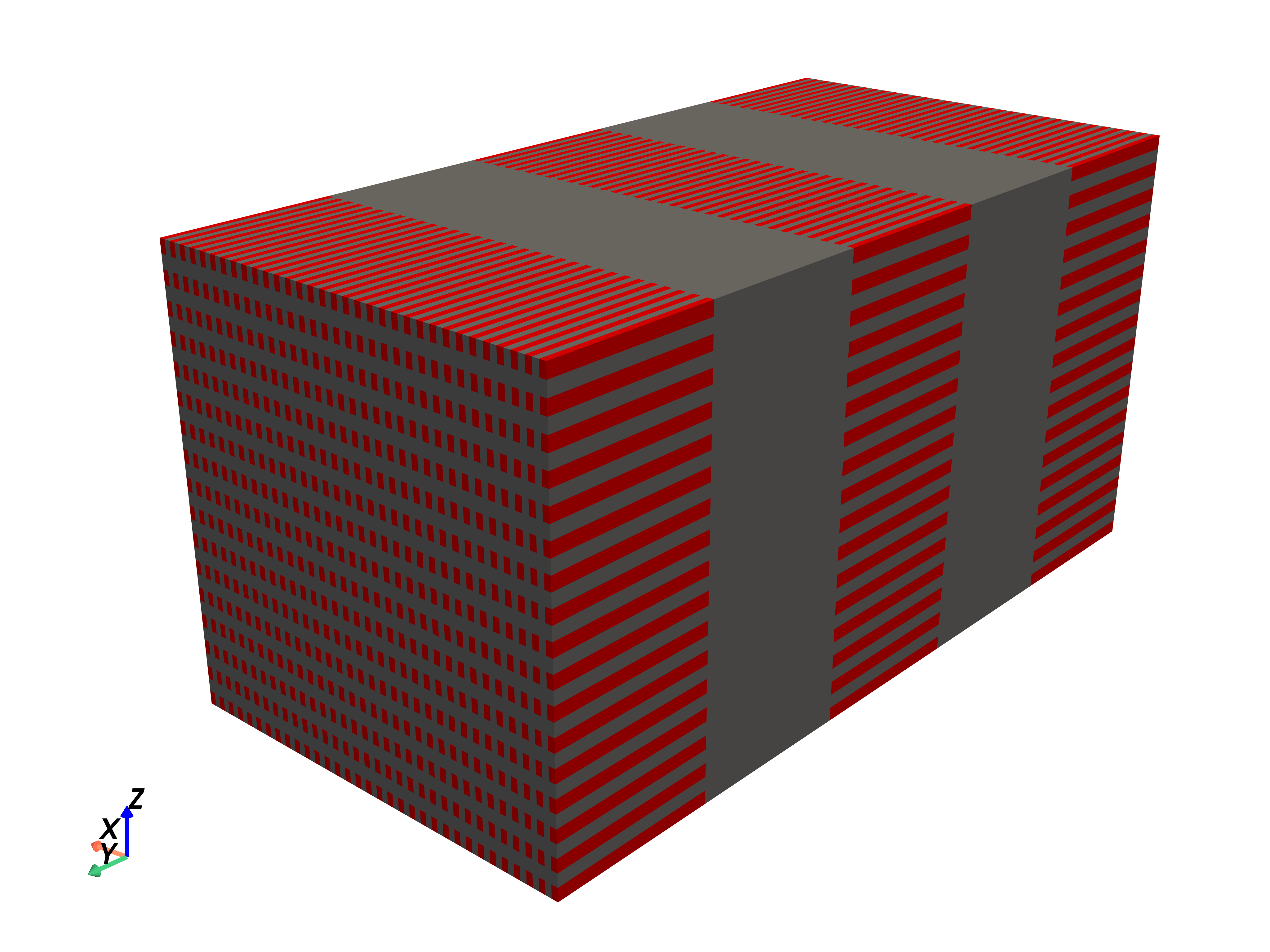}};
            \begin{scope}[x={(image.south east)},y={(image.north west)}]
                \spy[black, size=60] on (1.25,4.45) in node [left] at (2.9,6.1);
            \end{scope}
        \end{tikzpicture}
        \caption{Orthogonal layering directions}
        \label{fig:rank3_orth}   
    \end{subfigure}
    \begin{subfigure}[t]{0.49\textwidth}
        \centering
        \vskip 0pt
        \begin{tikzpicture}[spy using outlines={circle,yellow,magnification=5,size=1.5cm, connect spies}]
            \node[anchor=south west] (image) at (0,0) {\pgfimage[width=\textwidth]{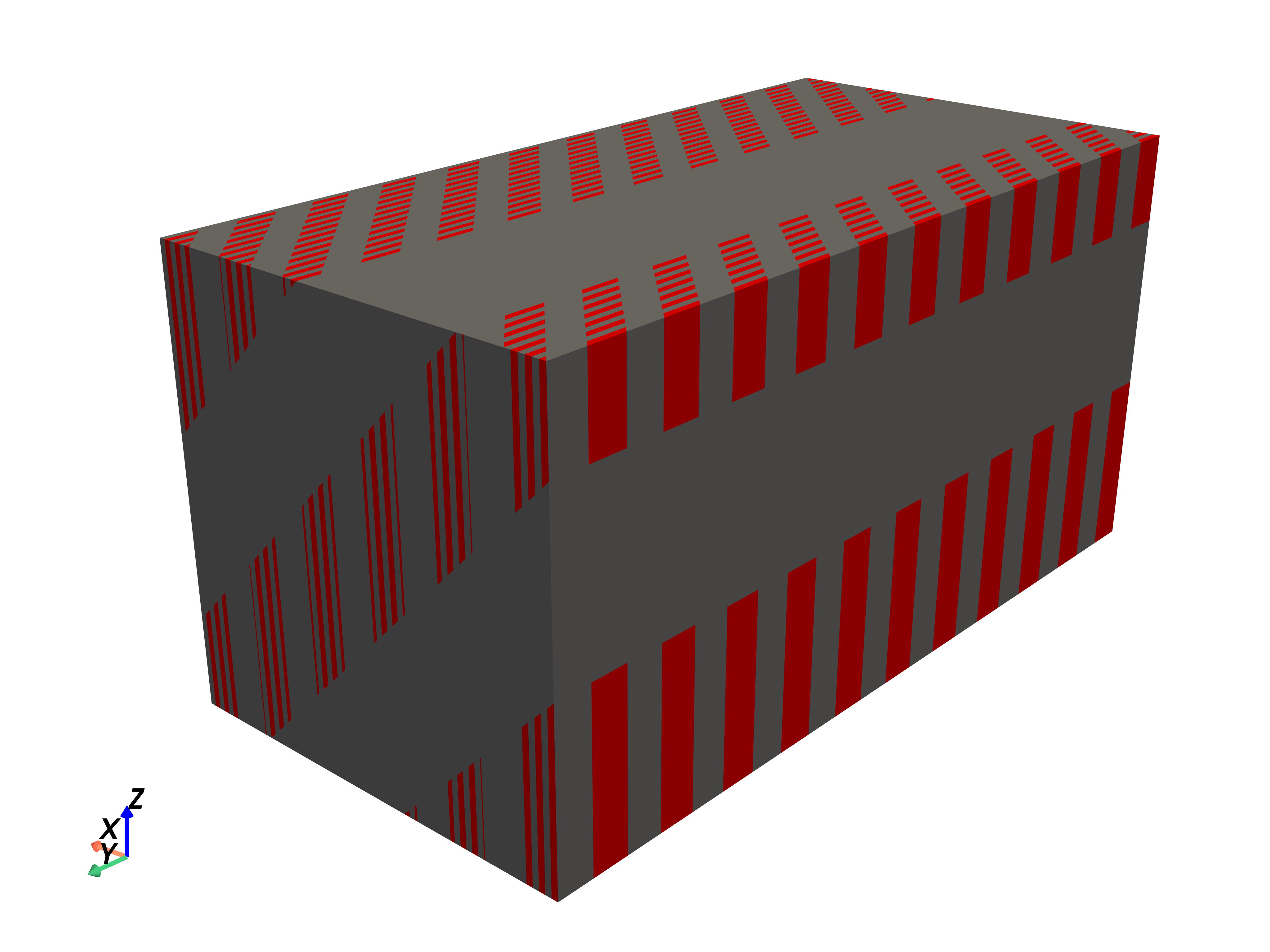}};
            \begin{scope}[x={(image.south east)},y={(image.north west)}]
                \spy[black, size=60] on (1.25,4.45) in node [left] at (2.9,6.1);
            \end{scope}
        \end{tikzpicture}
        \caption{Non-orthogonal layering directions}
        \label{fig:rank3_dmn}
    \end{subfigure}
    \caption{\fonems{R}ank-three, coated layered materials with orthogonal layering directions (a), and non-orthogonal layering directions (b).
    The core material is shown in red and the coating material is shown in gray. \fthrms{We hint
    at the separation of length scales between the material layers through our choice of the layer thicknesses.}}
    \label{fig:rank3_images}
\end{figure} As an example, two rank-three CLMs are depicted in Figure~\ref{fig:rank3_images}, where
the layering directions~$\fnr\in\setNorth$ of the material shown in Figure~\ref{fig:rank3_orth} are
\begin{equation}
    \setNorth = \left\{(1,0,0)_{\{\fe_i\}},(0,0,1)_{\{\fe_i\}},(0,1,0)_{\{\fe_i\}}\right\},
\end{equation}
and the layering directions~$\fnr\in\setnoNorth$ of the material shown in Figure~\ref{fig:rank3_dmn} are
\begin{equation}
    \setnoNorth = \left\{(1,0,0)_{\{\fe_i\}},\frac{1}{\sqrt{2}}\left(1,1,0\right)_{\{\fe_i\}},\frac{1}{\sqrt{2}}\left(1,0,1\right)_{\{\fe_i\}}\right\}.
\end{equation}
As described, the gray coating material~$m_2$ is repeatedly applied in layers,
such that for ${\mysf{r}}>1$, the $\mysf{r}$-th sub layer of the rank-three CLM is constructed by layering
a rank-(${\mysf{r}}-1$) layered material with the coating material~$m_2$. By definition,
the $0$-th sub layer of the rank-three CLM is purely made of the core material~$m_1$.
\ftwoms{We wish to compute the effective properties~$\Aeff$ of a coated layered material
by repeatedly applying the homogenization equation~\eqref{eq:Aeff_milton}.}
\ftwoms{For the recursive application of the rank-one
    homogenization \fonems{formula~\eqref{eq:Aeff_milton}} to be valid, the phases at
    each homogenization step must be homogeneous on the length scale of the respective layering step.
    This requires a separation of length scales, in the sense that
    the thickness~$\delta_{\mysf{r}}$ of the material sections in the rank-${\mysf{r}}$ sub layer is drastically smaller
    than the thickness~$\delta_{{\mysf{r}}+1}$ of the material sections in the next sub layer, i.e., $\delta_{\mysf{r}} \ll \delta_{\mysf{r}+1}$. 
    In Figure~\ref{fig:rank3_images}, we hint at this separation of length scales between the material layers 
    through our choice of the layer thicknesses. Nonetheless, we retain a visible degree of inhomogeneity,
    so that the layer structures can be distinguished for illustration purposes.}

\fonems{\ftwoms{Before we repeatedly apply equation~\eqref{eq:Aeff_milton} to derive an} expression for the effective properties~$\Aeff$, we first 
\fthrms{simplify} the expression~\eqref{eq:Aeff_milton} \fthrms{for the case at hand}.
We} let the reference material properties~\fonems{$\ffA_0$} in \fonems{equation~\eqref{eq:Aeff_milton}}
approach the properties~\fonems{$\ffA_2$} of the coating material~$m_2$\fonems{, resulting in the equation~\cite[§9]{milton2022theory} 
\fonems{
\begin{equation}\label{eq:Aeff_milton_limit}
    (1-f_2)(\Aeff-\ffA_2)^{-1} = \left(\ffA_1-\ffA_2\right)^{-1} + f_2\ftwoms{\GammaAtwo}(\fn).
\end{equation}
}Here, the coating material~$m_2$ must not be rigid or a void, \ftwoms{the tensor~$(\ffA_1-\ffA_2)$ must be invertible\fthrms{,
and~$f_2$ denotes the volume fraction of the coating material~$m_2$. Also,} the fourth order 
\ftwoms{tensor~$\GammaAtwo(\fnr) \in \spacefourth$ \fthrms{is defined by} equation~\eqref{eq:Gammadef} as}
\fonems{
\begin{equation}\label{eq:Gammadef_Atwo}
    \ftwoms{\GammaAtwo}(\fnr) = \GammaA(\fnr)\left[\GammaA(\fnr)\ffA_2\GammaA(\fnr)\right]^{\fthrms{\dagger}}\GammaA(\fnr),}
\end{equation}
where the inverse is taken on the subspace onto which~$\GammaA$ projects. }}\ftwoms{To
compute the effective properties~$\Aeff$ of a rank-$R$ CLM, we} recursively apply equation\fonems{~\eqref{eq:Aeff_milton_limit}} for the effective
material properties of rank-one layered materials,
such that~\cite[§9]{milton2022theory}
\begin{equation}\label{eq:rankthree_milton}
    \fonems{(1-f_2)(\Aeff - \ffA_2)^{-1} = (\ffA_1 - \ffA_2)^{-1} + f_2 \sum_{\mysf{r}=1}^{\fonems{R}} \csfr\ftwoms{\GammaAtwo}(\fnr).}
\end{equation}
\fonems{Here}\fthrms{,} the scalars~$\csfr$ are computed from the volume fractions~$f^{(\mysf{r})}_1$ of the core 
material~$m_1$ in the rank-${\mysf{r}}$ sub layer 
of the rank-three CLM, such that
\begin{equation}\label{eq:cis}
    \csfr = \frac{f^{({\mysf{r}}-1)}_1 - f^{({\mysf{r}})}_1}{f_2}\fthrms{,} 
\end{equation}
\fthrms{i.e., the properties
\begin{equation}
    \sum_{\mysf{r}=1}^{\fonems{R}}\csfr = 1 \ftwoms{\quad \text{and} \quad \csfr > 0,}
\end{equation}
hold.}
Because the rank-$0$ sub layer consists solely of the core material~$m_1$
by definition, the volume fraction~$f^{(0)}=1$ is one. \ftwotb{Equation~\eqref{eq:rankthree_milton}
is essential to \fthrms{compute the effective properties} of layered materials with infinite material contrast, and thus
for the treatment of such materials using a DMN architecture.}\ftwoms{ In
Appendix~\ref{sec:regularityconditions}, we employ equation~\eqref{eq:rankthree_milton} to
study the conditions under which the effective properties~$\Aeff$ of a 
CLM are singular when the core material~$m_1$ is \fthrms{rigid}. With the objective of 
applying FDMNs to suspensions of rigid fibers in mind, we summarize the results pertaining to rank-$R$ CLMs 
with a \fthrme{\textit{rigid}} core material~$m_1$
and an \fthrme{\textit{incompressible}} coating material~$m_2$ with the following statements. 
\fthrms{
\begin{enumerate}
    \item The effective properties~$\Aeff$ of a CLM of rank~$R<3$ are always singular. \label{statement:1}
    \item For rank-3 CLMs, the effective properties~$\Aeff$ are singular if at least one layering direction
    is orthogonal to two other layering directions, or if at least two layering directions are collinear.\label{statement:2}
    \item Rank-$3$ CLMs with mutually non-orthogonal and mutually non-collinear 
    layering directions~$\fn_1,\fn_2,$ and~$\fn_3$ have non-singular effective properties.
    Thus, in the context of this article, we require\label{statement:3}
    \begin{equation}\label{eq:nonmutual}
        0 < |\fn_1\cdot\fn_2| < 1 ,\quad 0<|\fn_1\cdot\fn_3|<1 ,\quad 0<|\fn_2\cdot\fn_3|<1.
    \end{equation}
\end{enumerate}
The statements~\ref{statement:1}-\ref{statement:3} are of critical importance for the treatment of incompressible
suspensions of rigid particles
using a DMN architecture. It follows that CLMs with an incompressible coating material and rigid core material
should be at least of rank~$R=3$ and have non-collinear and mutually non-orthogonal layering directions to be non-singular.
Therefore, we leverage such non-singular-singular rank-3 CLMs in the FDMN architecture to treat suspensions of rigid fibers,
as we explain in section~\ref{subsec:architecture_FDMN}.}
}
\section{Identifying deep material networks for suspensions of rigid \fthrme{fibers}}\label{sec:identifying_dmns}
\subsection{Architecture of direct deep material networks}\label{subsec:architecture_dDMN}
\foneme{\ftwoms{
Because the FDMN architecture \fthrms{extends} the (direct) DMN \fthrms{framework}, we briefly summarize
the direct DMN architecture in this section, before presenting the FDMN architecture in the next section~\ref{subsec:architecture_FDMN}. 
For in depth discussions and explanations of the direct DMN architecture, we refer 
the reader to articles by Gajek et al.~\cite{gajek2020micromechanics,gajek2021fe,gajek2022fe}.}
A two-phase direct DMN is defined as a perfect binary tree, consisting of two-phase 
laminates~$\lmb$ as nodes, see Figure~\ref{fig:DMN_weight_propagation}. 
The laminates~$\lmb$ are defined by a single lamination direction~$\fn_{\mysf{k}}^{\mysf{i}}$
and two volume fractions~$c_{\mysf{k},1}^1$ and~$c_{\mysf{k},1}^2$. 
The DMN tree is ordered, rooted and has depth~$K$. We use the letter~${\mysf{k}}=1,...,K$ to denote the depth of a node
and the letter~$\mysf{i}=1,...,2^{\mysf{k}-1}$ to indicate the horizontal position of a node in the respective layer.
\begin{figure}[h!]   
	\centering
	\begin{subfigure}[t]{.309\textwidth}  
		\includegraphics[width=\textwidth]{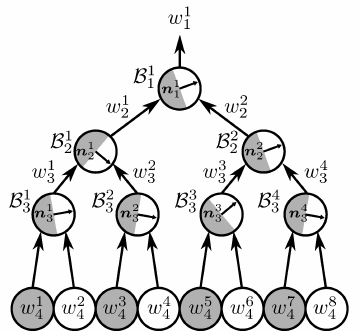}
		\caption{Weight propagation}
		\label{fig:DMN_weight_propagation} 
	\end{subfigure}
	\begin{subfigure}[t]{0.309\textwidth}
		\includegraphics[width=\textwidth]{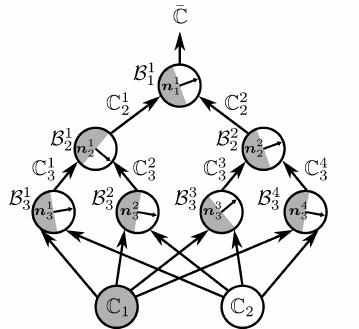}
		\caption{\ftwome{Stiffness} propagation}
		\label{fig:DMN_stiffness_propagation}  
	\end{subfigure}
	\caption{Weight and stiffness
    propagation (from the bottom to
    the top) in a two-phase direct
    DMN~\cite{gajek2021fe} of depth K = 3, by Gajek et al.~\cite[Fig. 1]{gajek2022fe},
    licensed under \href{https://creativecommons.org/licenses/by/4.0}{CC BY 4.0}}
	\label{fig:DMN_schematics}
\end{figure}
\ftwoms{We parametrize a direct DMN} by the two vectors~$\vecn\in(\reals^3)^{2^K-1}$ 
\begin{equation}
    \vecn = (\fn_{K}^1,....,\fn_{K}^{2^{K-1}},\fn_{K-1}^1,...,\fn_{K-1}^{2^{K-2}},...,\fn_1^1),
\end{equation}
and~$\vecw\in\reals^{2^K}$
\begin{equation}
    \vecw = (w_{K+1}^1,...,w_{K+1}^{2^K}).
\end{equation}
\ftwoms{Instead of the laminate volume fractions, 
the direct DMN is parametrized in terms of the weights~$w_{K+1}^{\mysf{i}}$}
to improve numerical stability during the training process~\cite{liu2019deep,liu2019exploring}.
The~$\mysf{i}$-th weight in the~$\mysf{k}$-th layer is computed from the pairwise summation of the weights in the previous 
layer~$\mysf{k}+1$, see Figure~\ref{fig:DMN_weight_propagation}, such that
\begin{equation}
    w_{\mysf{k}}^{\mysf{i}} = w_{\mysf{k+1}}^{2\mysf{i}-1} + w_{\mysf{k+1}}^{2\mysf{i}},
\end{equation}
and the volume fractions~$c_{\mysf{k},1}^{\mysf{i}}$ and~$c_{\mysf{k},2}^{\mysf{i}}$ follow from
\begin{equation}
    c_{\mysf{k},1}^{\mysf{i}} = \frac{w_{\mysf{k+1}}^{2\mysf{i}-1}}{w_{\mysf{k+1}}^{2\mysf{i}-1}+w_{\mysf{k+1}}^{2\mysf{i}}},
    \quad\text{and}\quad c_{\mysf{k},2}^{\mysf{i}} = 1-c_{\mysf{k},1}^{\mysf{i}}.
\end{equation}
Additionally, the weights~$w_{K+1}^{\mysf{i}}$ need to be positive and sum to unity, such that
\begin{equation}
    w_{K+1}^{\mysf{i}} \geq 0 \quad\text{and}\quad \sum_{\mysf{i}}^{2^K}w_{K+1}^{\mysf{i}} = 1.
\end{equation}\foneme{The vectors~$\vecn$ and~$\vecw$ fully parametrize a direct DMN of depth~$K$ and are identified by
fitting the linear homogenization function of the DMN to the linear homogenization function of the microstructure of interest.
The fitting process is carried out by using supervised machine learning, and we refer to this process as \textit{offline training}.
To evaluate the linear homogenization function of the DMN during the offline training,
input stiffness pairs are propagated through the \ftwome{network} of homogenization blocks~$\lmb$,
see Figure~\ref{fig:DMN_stiffness_propagation}. Consequently, if the stiffness~$\ffC_{\mysf{k}}^{\mysf{i}}$ of 
a homogenization block~$\lmb$ is singular, this singular stiffness can propagate through the whole DMN, which is undesirable 
for the offline training.
Once the parameter vectors~$\vecn$ and~$\vecw$ are known, they remain unchanged and are 
used in the~\textit{online evaluation} to predict the non-linear response of the studied material.
In sections~\ref{subsec:offline_training},~\ref{subsec:offline_training_fibers}, 
and~\ref{subsec:online_evaluation}, we discuss the offline training and the online evaluation for the proposed
FDMN architecture.
}
}
\subsection{\fthrme{Architecture of flexible deep material networks}}\label{subsec:architecture_FDMN}
\foneme{With the results of the previous \ftwome{sections~\ref{subsec:affinefields}-\ref{subsec:coated_layered_materials}} at hand,
we present an extended DMN architecture that is able to treat \fthrme{fluid phases and infinite material contrast}. 
Compared to the architecture of (direct) DMNs shown in Figure~\ref{fig:DMN_schematics},
the lowest layer of laminates is replaced by a layer of rank-$R$ CLMs~$\mathcal{C}_{R}$, see Figure~\ref{fig:FDMN_schematics}.
\begin{figure}[h!]   
	\centering
	\begin{subfigure}[t]{.589\textwidth}  
		\includegraphics[width=\textwidth]{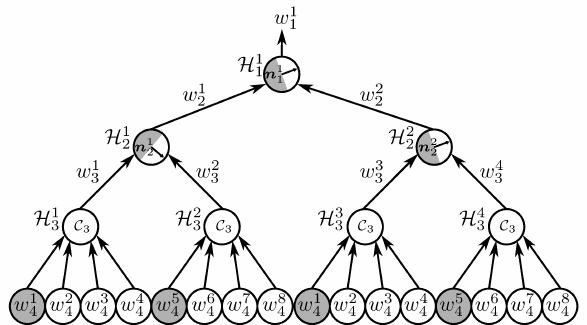}
		\caption{Weight propagation}
		\label{fig:FDMN_weight_propagation} 
	\end{subfigure}
	\begin{subfigure}[t]{0.309\textwidth}
		\includegraphics[width=\textwidth]{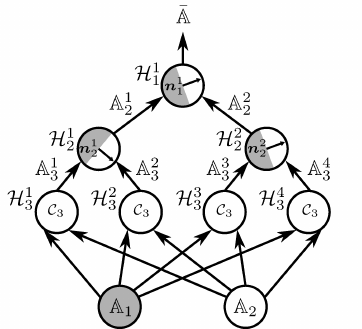}
		\caption{Material tensor propagation}
		\label{fig:FDMN_stiffness_propagation}  
	\end{subfigure}
	\caption{Weight and material property propagation (from the bottom to the top) in a two-phase \foneme{FDMN}~\cite{gajek2021fe} of depth $\depthDMN=3$
    \ftwome{with rank-$3$ CLMs}}
	\label{fig:FDMN_schematics}
\end{figure}The
rank~$R$ of the CLM~$\mathcal{C}_{R}$ 
and the restrictions on its layering directions~$\fnr$ are problem dependent, and chosen
such that the non-singularity condition~\eqref{eq:non-singularitycondition} is satisfied.
This guarantees that only non-singular material tensors~$\ffA_{\mysf{k}}^{\mysf{i}}$ are propagated through the network.
\fthrme{If all phases are kinematically unconstrained, solid, and non-singular, rank-$1$ CLMs can be employed to recover
the direct DMN architecture.
However, in the case of incompressible suspensions of rigid particles, the CLMs are of rank~$R=3$ and have mutually
non-collinear and non-orthogonal layering directions~$\fnr$, as per \mbox{statements~\ref{statement:1}-\ref{statement:3}}.}
\fthrme{Additionally, t}he homogenization blocks~$\lm$ implement the
general homogenization function defined by equation~\eqref{eq:Aeff_milton} in which the
projector function~$\GammaA$ depends on the physics of the investigated problem.
This allows the treatment of \fthrme{composites consisting of linear materials~$m\in\Mco$, including incompressible
linearly viscous fluids}.
\ftwoms{For} problems involving two-phase materials,
$R+1$ weights and two materials are assigned to each CLM.
The number of layering directions~$\Nnormals$
and the number of input weights~$\Nweights$ in a two-phase FDMN of depth~$\depthDMN$ with rank-$R$ CLMs are
\begin{equation}~\label{eq:Nnormals_Nweights}
\Nnormals = 2^{\depthDMN-1}(R+1)-1, \quad \Nweights = 2^{\depthDMN-1}(R+1).
\end{equation}
As an example, Figures~\ref{fig:FDMN_weight_propagation} and~\ref{fig:FDMN_stiffness_propagation} show
the weight and material property propagation of a two-phase FDMN of depth~$\fthrme{K=3}$ with two-phase CLMs of rank~$\fthrme{R=3}$ as bottom layer.
Of the four weights of a rank-$3$ CLM, one is associated with the core material~$m_1$, and three are associated with 
the coating material~$m_2$. This doubles the number of weights for an FDMN with rank-$3$ CLMs as compared to a DMN of the same 
depth~$\depthDMN$. Additionally, each rank-$3$ CLM block~$\clm$ has three layering directions~$\fnr$ instead of
one layering direction like rank-one laminates. \ftwoms{To reduce the number of
free parameters and ensure the CLMs are non-singular, the relative angles between the layering directions~$\fnr$
are fixed, while their joint orientation in space is determined through the offline training, as we explain
in section~\ref{subsec:offline_training} in more detail.}
Because CLMs are based on the repeated layering of rank-one layered materials,
an FDMN of depth~$\depthDMN$ can be considered to be a special case of a (direct) DMN of depth~$\depthDMN+(R-1)$ 
with generalized homogenization blocks. \foneme{Then, a CLM represents a subtree of the DMN of depth~$\depthDMN+(R-1)$,
in which the weights~$w_{\mysf{k}}^{\mysf{i}}$ of certain phases are set to zero, such that a CLM emerges, see Figure~\ref{fig:CLM_schematics}.
\begin{figure}[h!]   
	\centering
	\begin{subfigure}[t]{0.309\textwidth}  
		\includegraphics[width=\textwidth]{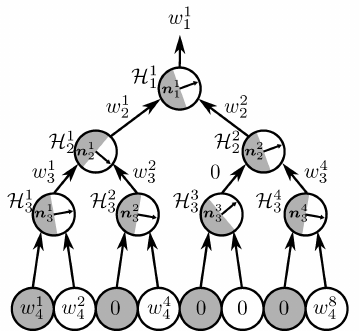}
		\caption{Weight propagation}
		\label{fig:CLM_weight_propagation} 
	\end{subfigure}
	\begin{subfigure}[t]{0.309\textwidth}
		\includegraphics[width=\textwidth]{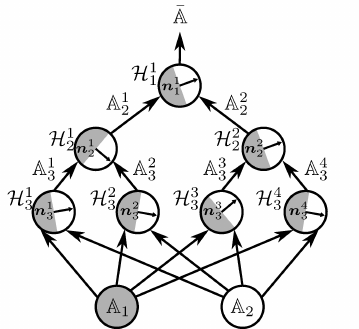}
		\caption{Material tensor propagation}
		\label{fig:CLM_stiffness_propagation}  
	\end{subfigure}
	\caption{Weight and material property propagation (from the bottom to the top) in a two-phase DMN of depth~$\fthrme{K=3}$,
    representing a two-phase rank-$3$ \foneme{CLM}}
	\label{fig:CLM_schematics}
\end{figure}
In particular, except for one phase, the weights associated with the singular core material~$m_1$ are set to
zero. Additionally,~$R$ weights associated with the non-singular coating material~$m_2$
must be non-zero, such that the rank~$R$ of the CLM is maintained.
Technically, (direct) DMNs are therefore able to treat infinite material contrast intrinsically.
However, the weights~$w_{\mysf{k}}^{\mysf{i}}$ need to be identified during the offline training, 
and might not always have appropriate values to generate CLMs with the required rank, such that singular
material tensors~$\ffA_{\mysf{k}}^{\mysf{i}}$ might propagate through the whole DMN.
Also, if the layering normals are not chosen appropriately, the singular material properties may propagate through the whole DMN as well.
Controlling the weights and normals via the implementation of the CLM blocks instead of learning them is therefore preferable to prevent 
singularities during the offline training. Furthermore, FDMNs of depth~$\depthDMN$ have fewer free parameters compared to
equivalent DMNs of depth~$\depthDMN+R-1$, which aids in the offline training as well.}

\foneme{Hypothetically, the presented FDMN architecture \fthrme{could} be applied to problems with more than two phases as well.
Let~$N_{\mysf{R}}$ denote the number of non-singular phases and~$N_{\mysf{S}}$ denote the number of singular phases.
Then, for each singular phase, the FDMN needs to contain at least one CLM to which the phase is assigned, and which has non-singular effective
properties. This requires a minimum of~$N_{\mysf{S}}$ CLMs. Additionally, the coatings of the CLMs can be constructed by mixing the 
non-singular phases, possibly by constructing~$N_{\mysf{R}}$-phase layered materials.
\ftwome{However, the application of FDMNs to materials with voids or more than two phases is beyond the scope of this article,
because we focus on suspensions of rigid fibers.}
}
}
\subsection{Material sampling}\label{subsec:material_sampling}
\foneme{
For materials with solid phases, DMNs are trained 
by sampling the linear elastic homogenization function of the microstructure of interest with randomly generated
input materials~\cite{liu2019deep,liu2019exploring,gajek2021fe,gajek2022fe}, and fitting the
DMN parameters to replicate the desired homogenization function. 
Similarly, in the case of rigid particles suspended in viscous media, we train \fthrme{FDMNs}
to learn the linear viscous homogenization function with the goal of 
predicting non-linear viscous behavior.
Consequently, we sample the linear viscous homogenization function
by using randomly generated\fthrme{,} \ftwoms{linearly viscous} input materials.}\foneme{
\ftwoms{We} base our sampling strategy on the approach used by Gajek et al.~\cite{gajek2021fe}.
In the article by Gajek et al.~\cite{gajek2021fe} on glass fiber reinforced polyamide composites,
the authors chose the stiffness samples \ftwoms{for} the fiber material to be isotropic,
and the stiffness samples \ftwoms{for} to the polyamide matrix to be transversely isotropic.
This \ftwoms{strategy leads} to a smaller sample space compared to sampling orthotropic stiffnesses for both the fiber and the matrix materials,
as suggested by Liu et al.~\cite{liu2019deep,liu2019exploring}.
To reduce the computational cost of sampling even further for the considered case of fibe\fthrme{r} suspension,
we only generate sample viscosities for the matrix material and consider the suspended fibers \ftwoms{to be} rigid in all computations.
To generate samples for the matrix material, we follow Gajek et al.~\cite{gajek2021fe} and
choose the sample viscosit\fthrtb{y}~$\Vsample$ \fthrtb{pertaining to the $s$-th sample} to be transversely isotropic, such that
\begin{equation}\label{eq:sampleviscos}
    \Vsample = 2\eta_2\left(\ffP_2 - g \fA'(\veca)\otimes\fA'(\veca)\right),
\end{equation}
where the scalar~$g\in[0,1)$ defines the magnitude of the perturbation by the unitary and deviatoric tensor~$\fA'(\veca)\in\symdev$,
the vector~$\veca\in\reals^4$ collects four angles, and we consider the scalar viscosities
\begin{equation}
    \eta_2 = 10^{p} \unit{GPa},\quad p\in[-3,3].
\end{equation}
\ftwoms{For a fiber reinforced polyamide, Gajek et al.~\cite{gajek2021fe} employed an \fthrms{ansatz}
of the form~\eqref{eq:sampleviscos} to sample the material tangents of a \mbox{J2-elastoplastic} matrix material.
Similarly, with the objective of applying FDMNs to fiber suspensions with a shear-thinning polyamide matrix
as detailed in section~\ref{subsec:material_description},
we use equation~\eqref{eq:sampleviscos} to sample the material tangents of a Cross-type~\eqref{eq:crosseq} matrix material.}
The tensor function~$\fA'$ is defined by an eigenvalue decomposition~\cite{gajek2021fe}
\begin{align}
    \fA':\reals^4&\rightarrow\symdev\fthrms{,}\\
    \T{(a_1,a_2,a_3,a_4)}&\mapsto\fQ(a_2,a_3,a_4)\fA(a_1)\T{\fQ(a_2,a_3,a_4)}\fthrms{,}
\end{align}
into the orthogonal tensor~$\fQ(a_2,a_3,a_4)$ and a diagonal tensor~$\fA(a_1)$ in terms of the vector~${\veca=\T{(a_1,a_2,a_3,a_4)}}$.
The tensor~$\fA(a_1)$ is constructed from the angle~$a_1$ \ftwoms{via~\cite{gajek2021fe}}\fthrtb{
\begin{align}
    \fA &= \frac {1}{\sqrt {{\frac {\xi_1+1}{2\,\xi_1 +1}}}} \sty{\mathrm{diag}} \left(\xi_2\cos\left(a_1\right), \xi_2\sin{\left(a_1\right)}, \frac{1}{\sqrt{2}}\right),\\
    \xi_1 &= \cos \left( a_{\mysf{1}} \right) \sin \left( a_{\mysf{1}} \right), \quad \xi_2 = \frac{-\sqrt{2}}{2 \cos{\left(a_1\right)} + 2 \sin{\left(a_1\right)}},
\end{align}}
and the tensor~$\fQ(a_2,a_3,a_4)$ represents a rotation around the direction~$\fn$, such that 
\begin{align}
    \fQ(a_2,a_3,a_4) &: \ffR^3 \rightarrow \ffR^3, \ \fx \mapsto \cos\left(a_2\right)\fx + \sin\left(a_2\right) \fn\times \fx + (1-\cos\left(a_2\right))(\fn\cdot \fx) \fn,\\
    \fn &= \sin\left(a_3\right) \cos\left(a_4\right)\fe_1 + \sin\left(a_3\right) \sin\left(a_4\right)\fe_2 + \cos\left(a_3\right)\fe_3.
\end{align}
Here, the operator~$\sty{\mathrm{diag}}$ constructs a second order tensor in the standard basis~$\fe_i$ of~$\reals^3$,
and the components of the vector~$\veca$ live on the intervals
\begin{equation}
    a_1 \in [0,2\pi],\quad a_2 - \sin(a_2) \in [0,\pi], \quad a_3\in[0,\pi], \quad a_4\in[0,2\pi].
\end{equation}
We sample the space of the variables~$p$ and $\veca$ using the scrambled Sobol sequence~\cite{owen1998scrambling,SOBOL196786},
and employ an FFT-based computational procedure to compute the resulting effective viscosities~$\Veffsample$~\cite{bertoti2021computational,sterr2023homogenizing},
see section~\eqref{sec:rigid_fibers_in_pa6}.
}
\\
\subsection{Offline training}\label{subsec:offline_training}
\foneme{Using the material sampling procedure described in the previous section~\ref{subsec:material_sampling},
we wish to optimize the parameters of the FDMN with regard to the training data set. For the case 
of rigid fibers suspended in viscous media, we consider FDMNs with rank-$3$ CLMs as bottom layer,
\ftwoms{which satisfy} the requirements derived in section~\ref{subsec:CLMs_ico_rigid}.
We collect all layering directions of the FDMN in a vector
\begin{equation}
    \vecn = (\fn_{K}^1,...,\fn_{K}^{3(2^{K-1})},\fn_{K-1}^1,...,\fn_{K-1}^{2^{K-2}},...,\fn_2^1,\fn_2^2,\fn_1^1),
\end{equation}
where we used the fact that the $k$-th layer contains~$2^{k-1}$ normals if~$k<K$, and the $K$-th layer contains~$3(2^{K-1})$ 
normals. Because the homogenization blocks in the $K$-th layer are rank-three CLMs, the number of normals is different
from the other layers.
Each normal~$\fn_{\mysf{k}}^{\mysf{i}}$ in the $k$-th layer for~$k<K$ is constructed using two angles~$a_{\mysf{k}}^{\mysf{i}}$
and~$b_{\mysf{k}}^{\mysf{i}}$ via
\begin{equation}
    \fn_{\mysf{k}}^{\mysf{i}} = \sin(b_{\mysf{k}}^{\mysf{i}})\cos(a_{\mysf{k}}^{\mysf{i}})\fe_1 +
    \sin(b_{\mysf{k}}^{\mysf{i}})\sin(a_{\mysf{k}}^{\mysf{i}})\fe_2 +
    \cos(b_{\mysf{k}}^{\mysf{i}})\fe_3.
\end{equation}
However, for the normals~$\fn_K^{\mysf{i}}$ in the $K$-th layer,
the procedure is different. To avoid singular effective properties during the evaluation of the FDMN in accordance
with the discussion in sections~\ref{subsec:coated_layered_materials} and~\ref{subsec:CLMs_ico_rigid}, 
we ensure that the three layering normals of a rank-three CLM block are not mutually non-orthogonal and
mutually non-collinear as follows. First, for each rank-three CLM block~$\clm^{\mysf{i}}$, 
we compute the components~$Q^{\mysf{i}}_{op}$ of the orthogonal tensor~$\fQ^{\mysf{i}}$ in the standard basis~$\{\fe_i\}$
using three \ftwoms{E}uler angles~$c_{\mysf{K}}^{\mysf{i}}, d_{\mysf{K}}^{\mysf{i}}, e_{\mysf{K}}^{\mysf{i}}\in[0,2\pi]$ via
\begin{equation}
    Q^{\mysf{i}}_{op} =
	\left[\begin{array}{ccc}
		1 & 0 & 0\\
		0 & \cos(c_{\mysf{K}}^{\mysf{i}}) & -\sin(c_{\mysf{K}}^{\mysf{i}})\\
		0 & \sin(c_{\mysf{K}}^{\mysf{i}}) & \cos(c_{\mysf{K}}^{\mysf{i}})\\
	\end{array}\right]
	\left[\begin{array}{ccc}
		\cos(d_{\mysf{K}}^{\mysf{i}}) & 0 & \sin(d_{\mysf{K}}^{\mysf{i}})\\
		0 & 1 & 0\\
		-\sin(d_{\mysf{K}}^{\mysf{i}}) & 0 & \cos(d_{\mysf{K}}^{\mysf{i}})\\
	\end{array}\right]
	\left[\begin{array}{ccc}
		\cos(e_{\mysf{K}}^{\mysf{i}}) & -\sin(e_{\mysf{K}}^{\mysf{i}}) & 0\\
		\sin(e_{\mysf{K}}^{\mysf{i}}) & \cos(e_{\mysf{K}}^{\mysf{i}}) & 0\\
		0 & 0 & 1\\
	\end{array}\right].
\end{equation}
Then, we compute the layering directions~$\fn_{\mysf{K}}^{\mysf{3(i-1)}+1},\fn_{\mysf{K}}^{\mysf{3(i-1)}+2},\fn_{\mysf{K}}^{\mysf{3(i-1)}+3}$ of the rank-three CLM~$\clm^{\mysf{i}}$ via
\begin{equation}
    \fn_{\mysf{K}}^{\mysf{3(i-1)}+1} = \fq^{\mysf{i}}_1,\quad \fn_{\mysf{K}}^{\mysf{3(i-1)}+2} = \frac{1}{\sqrt{2}}\left(\fq^{\mysf{i}}_1 + \fq^{\mysf{i}}_2\right),
    \quad \fn_{\mysf{K}}^{\mysf{3(i-1)}+3} = \frac{1}{\sqrt{2}}\left(\fq^{\mysf{i}}_1 + \fq^{\mysf{i}}_3\right),
\end{equation}
where~$\fq^{\mysf{i}}_1$, $\fq^{\mysf{i}}_2$, and $\fq^{\mysf{i}}_3$ are the first, second, and third column
vectors of the tensor~$\fQ^{\mysf{i}}$ in the standard basis~$\fe_i$, respectively.
This leads to a total of~$\Nangles$ angles with
\begin{equation}\label{eq:Nangles}
    \Nangles = 5\left(2^{K-1}\right) - 2.
\end{equation}
\ftwoms{The angles between the layering directions are fixed for each CLM~$\clm$, but the three layering
directions are rotated together during training.} Other procedures than the one
presented here are viable as well, as long as the constructed
vectors~$\fn_{\mysf{K}}^{\mysf{3(i-1)}+1}$, $\fn_{\mysf{K}}^{\mysf{3(i-1)}+2}$, and~$\fn_{\mysf{K}}^{\mysf{3(i-1)}+3}$
are mutually non-collinear and non-orthogonal. 
\foneme{Alternatively, one could parametrize each of the
three layering directions \linebreak$\fn_{\mysf{K}}^{\mysf{3(i-1)}+1}$,~$\fn_{\mysf{K}}^{\mysf{3(i-1)}+2}$, and~$\fn_{\mysf{K}}^{\mysf{3(i-1)}+3}$ with two angles each,
and identify the angles independently. Then, 
the effective properties of the CLM~$\clm^{\mysf{i}}$ are non-singular almost surely, because
the probability of any two of the three layering directions being collinear or one vector being
orthogonal to the other two is zero. However, parametrizing each of the three layering directions 
with two angles requires twice the amount of angles in the CLM layer of the FDMN as compared
to the presented procedure.}

To define the linear homogenization function of the FDMN, we collect all~$\Nangles$ angles of the layering directions in the vector~$\vecalpha$
\begin{equation}
    \vecalpha = (c_{\mysf{K}}^{1},d_{\mysf{K}}^{1},e_{\mysf{K}}^{1},...,c_{\mysf{K}}^{3(2^{K-1})},d_{\mysf{K}}^{3(2^{K-1})},e_{\mysf{K}}^{3(2^{K-1})},
                a_{\mysf{K-1}}^1,b_{\mysf{K-1}}^1,...,a_{\mysf{K-1}}^{2^{K-2}},b_{\mysf{K-1}}^{2^{K-2}},...,a_1^1,b_1^1),
\end{equation}
and collect all~$\Nweights$ input weights of the FDMN in the vector
\begin{equation}
    \vecw = (w_{K+1}^1,...,w_{K+1}^{\Nweights}).
\end{equation}
We follow Gajek et al.~\cite{gajek2020micromechanics} and define the loss 
function~$\lossdmn:\reals^{\otimes\Nangles}\times\reals^{\otimes\Nweights}\rightarrow\reals$ 
in terms of the linear homogenization 
function~$\DMNLIN:\mathcal{A}\times\mathcal{A}\times{\reals}^{\otimes\Nangles}\times{\reals}^{\otimes\Nweights}\rightarrow\mathcal{A}$ of the FDMN
as
\begin{equation}\label{eq:lossdmn}
	\lossdmn\left(\vecalpha, \vecw\right) = \frac{1}{\Nbatch} \sqrt[\ffoutb{q}]{\sum_{s=1}^{\Nbatch} \left(\frac{\norm{\Aeff^s - \DMNLIN\left(\ffA^s_{1}, \ffA^s_{2}, \vecalpha, \vecw\right)}_{\ffoutb{p}}}{\norm{ \Aeff^s }_{\ffoutb{p}}}\right)^{\ffoutb{q}}} 
    + \lambda \left( \sum_{i=1}^{\Nweights} w_{K+1}^i  - 1\right)^2,
\end{equation}
where~$\Nbatch$ is the number of samples,~$\ffA_1^s,\ffA_2^s\in\mathcal{A}$ denote the material tensors of the phases of a sample, and~$\Aeff^s\in\mathcal{A}$ denotes the effective material tensor
of a sample. Additionally,\fthrms{ the class of material tensors}\ftwoms{~$\mathcal{A}\in\{\symfour,\symdevfour\}$ depends on the considered problem},
\ffoutb{we follow Gajek et al.~\cite{gajek2020micromechanics} and choose the
constants~$p,q\in\naturals_{\geq 0}$ as~$p=1$ and~$q=10$,} and the operator~$\norm*{\cdot}_1$ \ftwoms{refers to} the $\ell^1$-norm of the components in Mandel notation.
The second summand in equation~\eqref{eq:lossdmn} encodes the simplex constraint on the weights of the DMN
\begin{equation}
    \sum_{\mysf{i}=1}^{\Nweights} w_{K+1}^{\mysf{i}} = 1,\quad\text{where}\quad w_{K+1}^{\mysf{i}} \geq 0,
\end{equation}
in the form of a quadratic penalty term with the penalty parameter~$\lambda>0$.
We implement the offline training in the Python programming language and use the 
machine learning framework PyTorch~\cite{pytorch} to identify the vectors~$\vecalpha$ and~$\vecw$. 
To do so, we leverage the automatic differentiation capabilities of PyTorch to solve the minimization problem
\begin{equation}
    \lossdmn\left(\vecalpha, \vecw\right) \rightarrow \min_{\vecalpha, \vecw},
\end{equation}
using the~RAdam\cite{liu2019variance} optimizer. We use mini batches of~$\Nbatch$ samples to compute the parameter updates
\begin{equation}
    \vecalpha_{q+1} = \vecalpha_q - \kappa_{q,\alpha} {\pdv{\lossdmn}{\vecalpha}}(\vecalpha,\vecw), \quad
    \vecw_{q+1} = \vecw_q - \kappa_{q,w} {\pdv{\lossdmn}{\vecw}}(\vecalpha,\vecw)
\end{equation}
for every epoch~$q$. Also, we follow previous work~\cite{liu2019deep,liu2019exploring,gajek2020micromechanics} and use perfect binary trees 
without tree compression~\cite{liu2019deep,liu2019exploring,gajek2021fe}
}
\subsection{Online evaluation}\label{subsec:online_evaluation}
\ftwome{In section~\ref{subsec:affinefields}, we showed that the velocity fields in layered emulsions are phase-wise affine,
and thus the online evaluation procedure of direct DMNs can be employed for FDMNs as well if the phases are kinematically unconstrained.}
\foneme{However, for FDMNs with incompressible phases, the online evaluation procedure needs to be adapted, as we discuss in the following.
We define the linear operator $\matA: \spacejumps \rightarrow \spacefluct$ which maps
the emulsion-wise jumps~$\vecb\in\spacejumps$ of the velocity gradient onto the phase-wise strain rate fluctuations.
\ftwoms{Also, we express the} effective dissipation potential density~$\Psieff:\spacefluct\rightarrow\reals$ of the FDMN in terms of the phase-wise
dissipation potentials~$\Psi_{\mysf{i}}:\symdev\rightarrow\reals$, the respective weights~$w_{K+1}^{\mysf{i}}$, and the phase-wise 
strain rate tensor~$\fD_{\mysf{i}}\in\symdev$ \ftwoms{as}
\begin{equation}
    \Psieff\psieffargs = \sum_{i=1}^{\Nweights}w_{K+1}^{\mysf{i}}\Psi_{\mysf{i}}(\fD_{\mysf{i}}),
\end{equation}
where all entries of the vector~$\vecDeff\in\spacefluct$ equal the effective strain rate~$\Deff$, i.e., 
\begin{equation}
    \vecDeff = (\Deff, \Deff, ..., \Deff).
\end{equation}
Thus, the minimization problem for the online, non-linear evaluation of the FDMN is
\begin{equation}\label{eq:onlineproblem}
    \Psieff\psieffargs \rightarrow \min_{\vecb \in \spacejumps},
\end{equation}
and the Euler--Lagrange equation of the problem~\eqref{eq:onlineproblem} reads
\begin{equation}\label{eq:eulerlagrange}
    \pdv{\Psieff}{\vecb}\psieffargs = \T{\matA}\pdv{\Psieff}{\vecD}\psieffargs = \veczero.
\end{equation}
In terms of a weight matrix~$\matW: \spacefluct \rightarrow \spacefluct$ defined \ftwoms{by} its action on a vector~${\vecd\in\spacefluct}$
\begin{equation}
    \matW \, \vecd = \left(w_{K+1}^1d_1, w_{K+1}^2d_2, ..., w_{K+1}^{\Nweights}d_K\right),
\end{equation}
we may write the Euler--Lagrange equation~\eqref{eq:eulerlagrange} as
\begin{equation}\label{eq:eulerlagrangesigma}
    \T{\matA}\matW\,\vectau\psieffargs = \veczero,
\end{equation}
with the vector of phase wise stresses~$\vectau\psieffargs\in(\symdev)^\Nweights$.
Therefore, the jump vector~$\vecb$ satisfying equation~\eqref{eq:eulerlagrangesigma} \ftwoms{is} determined using a Newton scheme 
with the update rule
\begin{equation}
    \vecb_{n+1} = \vecb_n + s_n\Delta\vecb_n,
\end{equation}
containing the iteration count~$n$, the backtracking factor~$s_n$, and the update direction~$\Delta\vecb_n$.
The update direction~$\Delta\vecb_n$ \ftwoms{solves} the linear system
\begin{equation}\label{eq:updaterule}
    \matH\Delta\vecb_n = \left(\T{\matA}\matW\,\vectau\psieffargs\matA\right)\Delta\vecb_n = \T{\matA}\matW\,\vectau\psieffargs.
\end{equation}
So far, the procedure is completely analogous to the approaches for (direct) DMNs presented in previous 
work~\cite{gajek2020micromechanics,gajek2021fe,gajek2022fe}.
However, in the case of incompressible phases, the matrix~$\matA$ is symmetric but singular,
\ftwoms{and the solution of the equation system~$\eqref{eq:updaterule}$
is not unique}. However, because of incompressibility, we know that the $\mysf{i}$-th entry~$\fb_{\mysf{i}}$ of the jump vector~$\vecb$ 
is orthogonal to the $\mysf{i}$-th entry~$\fn_{\mysf{i}}$ of the vector of layering normals~$\vecn$\ftwoms{, i.e.,
\begin{equation}
    \fb_{\mysf{i}}\cdot\fn_{\mysf{i}} = 0.
\end{equation}}Hence, we construct the block diagonal matrix~$\matN:\spacejumps \rightarrow \spacejumps$ consisting of~$\Nweights$ 
blocks~$\fN_{\mysf{i}}$ on the main diagonal, which are defined as
\begin{equation}
    \fN_{\mysf{i}} = \beta_{\mysf{i}}\fn_{\mysf{i}}\otimes \fn_{\mysf{i}},\quad \mysf{i}\in\{1,...,\Nweights\}
\end{equation}
with the constant~$\beta_{\mysf{i}}\in\reals_{\ftwoms{>0}}$.
By definition \ftwoms{it holds that
\begin{equation}
    \fN_{\mysf{i}}\fb_{\mysf{i}} = \fzero,
\end{equation}
and} the jump vector~$\vecb$ is in the kernel of~$\matN$, such that 
\begin{equation}\label{eq:nullspace}
    \matN\,\vecb = \veczero.
\end{equation}
Thus, we add equations~\eqref{eq:updaterule} and~\eqref{eq:nullspace}, such that the \ftwoms{solution~\fthrme{$\Delta\vecb_n$} of the resulting update rule}
\begin{equation}\label{eq:updaterule_ico}
    \left(\matH + \matN\right)\Delta\vecb_n = \T{\matA}\matW\,\vectau\psieffargs,
\end{equation}
is \ftwoms{unique}.
However, floating point precision may cause 
issues during the addition of the operators~$\matH$ and~$\matN$, if numbers of largely different magnitudes are summed.
To remedy this issue, the scalars~$\beta_{i}$ need to be chosen according to the magnitude of \foneme{the} entries
in the matrix~$\matH$. 
\ftwoms{Furthermore, some input weights~$w_{K+1}^{\mysf{i}}$ might become equal to, or close to zero during training,
which might render the system matrix~\fthrme{$(\matH+\matN)$} ill-conditioned.}
\ftwoms{Thus, it} might be difficult to obtain accurate update directions~\fthrme{$\Delta\vecb_n$} by solving equation~\eqref{eq:updaterule_ico}.
\ftwoms{While it is generally possible to prune the FDMN tree~\cite{liu2019exploring} by removing
phases with vanishing or almost vanishing volume fractions, this might collapse the rank-$3$ CLMs of the bottom
FDMN layer into layered materials of a lesser rank. In case of infinite material contrast,
this could result in the propagation of singular effective properties through the
FDMN. It is therefore preferable to work with possibly ill-conditioned systems and to employ
appropriate methods for the solutio\fthrme{n} of ill-conditioned systems~\cite{neumaier1998solving}.}
Alternatively, multiple FDMNs can be trained
and the ones with the best online evaluation results can be selected. In this article, we follow the latter approach.}
\section{\ftwoms{Application to rigid} fibers suspended in polyamide 6}\label{sec:rigid_fibers_in_pa6}
\subsection{Material description and computational aspects}\label{subsec:material_description}
\foneme{
Let a fiber orientation state be defined by the fiber orientation distribution function~$\rho:\twosphere\rightarrow\reals$,
which encodes the probability that fibers are oriented in the direction~$\fp\in\twosphere$.
In component scale molding simulations, \ftwoms{the orientation distribution function~$\rho$
is often spatially inhomogeneous, and it is therefore} prohibitively expensive to
compute the \ftwoms{spatial and temporal} evolution of \ftwoms{the function}~$\rho$~\cite{kennedy2013flow,wang2018molding}.
Therefore, the evolution of surrogate quantities that contain some of the information encoded in the
function~$\rho$ is often considered instead.
One such quantity is the second order fiber orientation tensor~\cite{ken1984distribution,advani1987use}
\begin{equation}
    \fN = \int_{\twosphere}\fp\otimes\fp \, \rho(\fn) \dif S(\fn),
\end{equation}
which is commonly used in component scale molding simulations
to track the evolution of the fiber orientation state~\cite{kennedy2013flow,wang2018molding}.
\begin{figure}[h!]
    \centering 
    \begin{subfigure}[t]{0.49\textwidth}
        \centering
        \vskip 0pt
        \includegraphics[trim=0 0 0 0,clip]{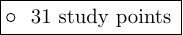}
        \includegraphics[trim=0 0 0 0,clip,width=\textwidth]{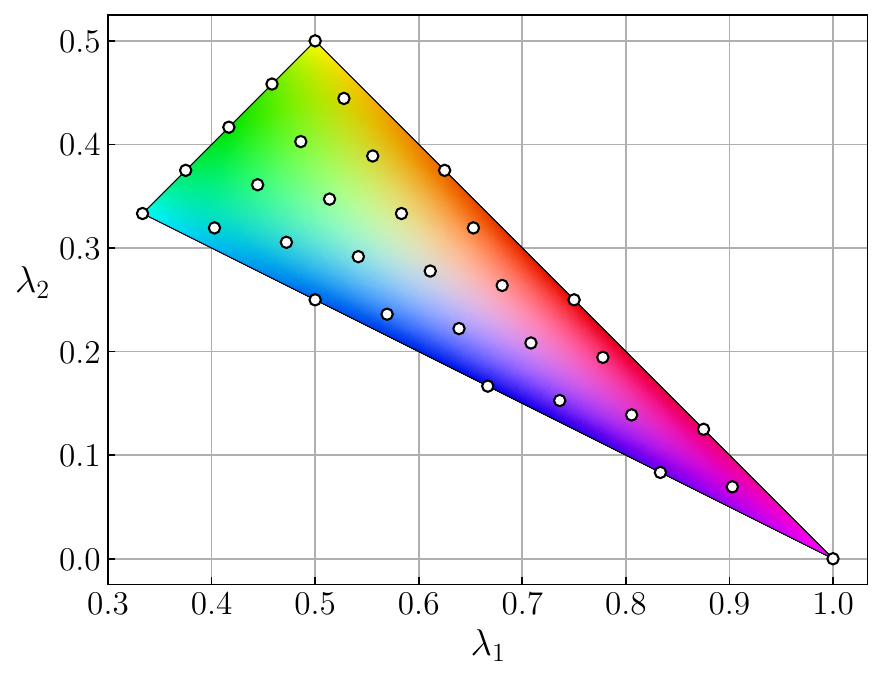}
        \vskip 1pt
        \caption{Fiber orientation triangle~$\setFOT$}
        \label{fig:FOT_CMYK}
    \end{subfigure}
    \begin{subfigure}[t]{0.49\textwidth}
        \centering
        \vskip 0pt
        \includegraphics[trim=0 165.5 0 0,clip,width=\textwidth]{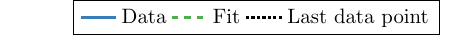} 
        \vskip 2.4pt
        \includegraphics[trim=0 2 0 17.5,clip, width=\textwidth]{eta_cross.pdf}  
        \caption{Cross-type fit}
        \label{fig:etamatrix}
    \end{subfigure}
    \caption{Fiber orientation triangle~{$\setFOT$} in CMYK coloring with 31 evaluation points (a), and material data with
    Cross-type fit for Ultramid{\sffamily\textregistered}B3K (b).}
    \label{fig:FOT_and_cross}
\end{figure}
Because a major prospect of FDMNs is the combination of component scale molding simulations 
with high fidelity information from full field viscous computations, we employ the orientation tensor~$\fN$ to describe
the fiber orientation state of the considered microstructures.} \foneme{The
tensor~$\fN$ may be parametrized by its two \ftwoms{largest} eigenvalues~$\lambda_1$ and~$\lambda_2$, 
as well as a \fthrtb{proper orthogonal tensor}~$\fR$ via~\cite{cintra1995orthotropic,bauer2022variety}
\begin{equation}
    \fN = \fR\, \sty{\mathrm{diag}}(\lambda_1,\lambda_2,1-\lambda_1-\lambda_2)\,\T{\fR},\quad \lambda_1\geq\lambda_2.
\end{equation}
\ftwoms{By objectivity~\cite{sterr2024machine}}, and because the tensor~$\fN$ has unit trace and is positive semi-definite, 
\ftwoms{every fiber orientation tensor~$\fN$ represents} a point in the 
fiber orientation triangle~$\setFOT$~\cite{kobler2018fiber,bauer2022variety}
\begin{equation}\label{eq:FOTriangle}
    \setFOT = \left\{\FOT = \T{(\lambda_1,\lambda_2)} \quad\middle|\quad 1 \geq \lambda_1 \geq \frac{1}{3} \quad \textrm{and} \quad \min\left(\lambda_{1}, 1 - \lambda_{1}\right) \geq \lambda_2 \geq \frac{1-\lambda_{1}}{2}  \right\}.
\end{equation}For detailed discussions regarding the description of fiber orientation states, we refer the reader to
Kanatani~\cite{ken1984distribution} and Bauer et al.~\cite{bauer2022variety, bauer2023phase}.
To generate the training data and the online validation data for the FDMN,
we use FFT-based computational procedures to compute the effective viscous response
of fiber suspensions~\cite{bertoti2021computational,sterr2023homogenizing}.
We generate the required fiber suspension microstructures using the sequential addition
and migration method~\cite{schneider2017sequential} for 31 orientation states~$\FOT$ in
the fiber orientation triangle~$\setFOT$, see Figure~\ref{fig:DOT}. For the visualization
of the orientation triangle~$\setFOT$, we follow K\"obler et al.~\cite{kobler2018fiber}, and use a CMYK coloring scheme.
To keep the computational effort for computing the effective viscous responses feasible,
we restrict to microstructures with a fiber volume fraction of 25\%, and assign all fibers
the length~$\ell$ and the diameter~$d$. Also, we set the aspect ratio~$r_{\mysf{a}}=\ell/d$ of all fibers to 10,
and use the results of Sterr et al.~\cite{sterr2023homogenizing} regarding the required resolution and size of the 
microstructure volume elements. Consequently, we choose a resolution of 15 voxels per fiber diameter, as well as cubic 
volume elements with edge size~$L=2.2\ell$.
Regarding the matrix material, we build on the investigations in Sterr et al.~\cite{sterr2023homogenizing,sterr2024machine},
and prescribe a commercially available polyamide 6~\cite{PA6data}
as the matrix material. We fit a Cross\fthrme{-}type \fthrtb{constitutive equation}
\begin{equation}\label{eq:crosseq}
	\eta(\gdot) = \eta_{\infty} + \frac{\eta_0-\eta_{\infty}}{1 + (k\gdot)^m},
\end{equation}
to the available material data in the interval~${[1.7, 16300] \persec}$ at a temperature of ${250^\circ\unit{C}}$,
and collect the resulting model parameters in Table~\ref{tab:crossparams}.
\begin{table}[h!]
    \centering
    \begin{tabular}{cccc}
        \hline
        $\eta_0$ & $\eta_{\infty}$ & $k$ & $m$ \\
        \hline
        288.9 $\Pas$ & 15.0 $\Pas$ & $10.9\cdot10^{-4}$ \ffoutb{$\unit{s}$} & 1.1\\
        \hline
    \end{tabular}
    \caption{{Parameters of the Cross-type \fthrtb{constitutive equation}~\eqref{eq:crosseq} for a commercially available polyamide 6~\cite{PA6data}}.}
    \label{tab:crossparams} 
\end{table}
For all non-Newtonian considerations, we {investigate} the six load cases collected in the matrix~$\Dmatrix$ in Mandel notation
\begin{equation}\label{eq:Ds}
	\Dmatrix = \gdot\sqrt{\frac{2}{3}}
	\left[\begin{array}{rrrrrr}
		1 &  -\frac{1}{2} &  -\frac{1}{2} & 0 & 0 & 0\\
		-\frac{1}{2} & 1 &  -\frac{1}{2} & 0 & 0 & 0\\
		-\frac{1}{2} &  -\frac{1}{2} & 1 & 0 & 0 & 0\\
		0 &  0 &  0 & \sqrt{\frac{3}{2}} & 0 & 0\\
		0 &  0 &  0 & 0 & \sqrt{\frac{3}{2}} & 0\\
		0 &  0 &  0 & 0 & 0 & \sqrt{\frac{3}{2}}
	\end{array}\right],
\end{equation}or each macroscopic scalar shear rate~$\gdot$ in the set of studied shear rates~$\setgdot$, such that
\begin{equation}\label{eq:gammadots}
    \gdot \in \setgdot = \gammaset.
\end{equation}
Consequently, the set~$\setD$ of all investigated load cases is defined by the \ftwoms{equations~\eqref{eq:Ds} and~\eqref{eq:gammadots}.}
\ftwoms{Because the viscous stress inside the rigid fibers is, constitutively, not well-defined~\cite{sterr2023homogenizing},
we employ a dual formulation of the associated homogenization problem for the FFT-based computations.
We discretize the microstructures on a staggered grid~\cite{harlow1965numerical},
and solve} the resulting equation systems with the conjugate gradient method \fthrme{(}CG) for linear computations,
and with a Newton-CG approach for non-linear computations.
With the goal of training an FDMN for each of the 31 fiber orientation states shown in Figure~\ref{fig:FOT_CMYK},
we generate 32 sample viscosities with the procedure described in section~\ref{subsec:material_sampling},
and compute the corresponding effective viscosities. To compute a single effective viscosity,
five FFT-based computations are necessary~\cite{bertoti2021computational},
leading to a total of 4960 computations, and 992 computed effective viscosities.}\foneme{ In
the previous sections~\ref{subsec:coated_layered_materials} and~\ref{subsec:architecture_FDMN}, we studied the linear homogenization functions of CLMs, 
and presented the FDMN architecture to treat infinite material contrast and incompressible materials. In the following,
we discuss the material sampling for the offline training,
the offline training procedure, and the online evaluation of FDMNs for suspensions of rigid particles.}
\subsection{Offline training}\label{subsec:offline_training_fibers}
\foneme{With the linear training data at hand, we wish to train FDMNs for the prediction
of the non-Newtonian viscous behavior of shear-thinning fiber polymer suspensions.
We follow previous \ftwoms{work}~\cite{gajek2020micromechanics,gajek2021fe} and choose the depth of the FDMN as~$K=8$ to achieve sufficient 
prediction quality for non-linear computations. An FDMN of depth~$K=8$ with CLMs of rank~$\fthrme{R=3}$ has 512 weights and 638 angles as free parameters,
as per equations~\eqref{eq:Nnormals_Nweights} and~\eqref{eq:Nangles}.
\foneme{\ftwome{For} each microstructure, we define the training data~$\setdata$ as
\begin{equation}
    \setdata = \{(\Vsample_1,\ffV_{\mysf{F}}\ftwoms{,}\Veffsample) \quad | \quad \mysf{s} \in (1,...,32)\},
\end{equation}
where~$\Vsample_1$ denotes the sample viscosities generated with equation~\eqref{eq:sampleviscos},
~$\ffV_{\mysf{F}}$ denotes the infinite viscosity of the rigid fibers, and~$\Veffsample$ stands for the effective viscosity of the sample~$\mysf{s}$.
}For each microstructure, we split the training data~$\setdata$
into the training set~$\settrain\subset\setdata$ and the validation set~$\setvalid\subset\setdata$, which consist of 90\% and 10\% of the total training data, respectively. 
The training set~$\settrain$ and the validation set~$\setvalid$ share no samples, i.e.,~$\settrain\cap\setvalid=\emptyset$.

As an initial guess for the FDMN parameters,
we uniformly sample all angles~$\vecalpha$ and weights~$\vecw$ from their respective intervals, and rescale the weights~$\vecw$ such that they sum
to unity. Because we use a relatively small set of 32 samples as training data, we dedicate a large portion of it to
the training set~$\settrain$, and choose to train 20 FDMNs per fiber orientation state~$\FOT$.
Furthermore, we train the FDMN\ftwome{s} on mini batches with size~$\Nbatch=8$, for which
we draw randomly from the training set~$\settrain$.
In case the last batch is smaller than eight samples, we drop the batch.
\foneme{Because the initialization of the parameter vectors~$\vecalpha$ and~$\vecw$, as well as the
offline training process are random, the FDMNs differ in their parameters and quality of fit.
By training 20 FDMNs per fiber orientation state, we leverage this randomness with the goal of obtaining FDMNs with a high
quality of fit. This strategy aims to reduce the total computational effort required to obtain sufficiently accurate FDMNs, 
because repeatedly training FDMNs on 32 samples requires less computational resources than conducting full field simulations
for larger sample sizes.} For
the minimization of the loss function~$\lossdmn$, we choose the RAdam algorithm~\cite{liu2019variance}
and use a learning rate sweep~\cite{smith2019super} to determine the learning rates~$\kappa_{0,\alpha}$ and $\kappa_{0,\mysf{w}}$.
With the learning rate sweep we obtain highly similar learning rates for both parameter groups~$\vecalpha$ and~$\vecw$,
such that~${\kappa_{0,\alpha} = \kappa_{0,\mysf{w}} = 1\cdot10^{-2}}$.
As the learning process \ftwoms{advances}, we use PyTorch's StepLR learning rate scheduler to 
improve convergence towards minima~\cite{darken1990note,darken1992learning}. To do so, we multiply the learning rates~$\kappa_{0,\alpha}$
and~$\kappa_{0,\mysf{w}}$ with a constant factor~${f_{\mysf{LR}}=0.75}$ every 150 epochs. Other than the learning rates,
we used standard hyperparameters for the two momentum coefficients~$\beta_1=0.9$ and~$\beta=0.999$, as well as 
the stabilization constant~$\varepsilon=10^{-8}$. Overall, we train every FDMNs for a total of 2000 epochs each.
}
\begin{figure}[H]
    \centering 
    \begin{subfigure}{0.49\textwidth}
        \centering
        \includegraphics[width=\textwidth]{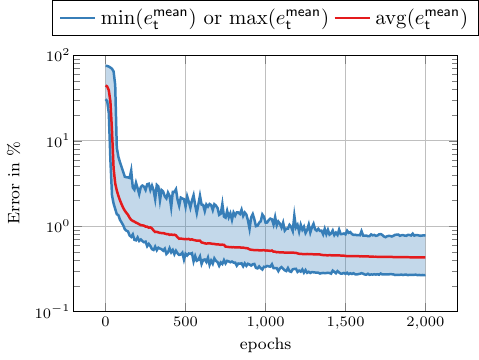}
        \caption{Training set~$\settrain$}
        \label{fig:train_mean}
    \end{subfigure}
    \begin{subfigure}{0.49\textwidth}
        \includegraphics[width=\textwidth]{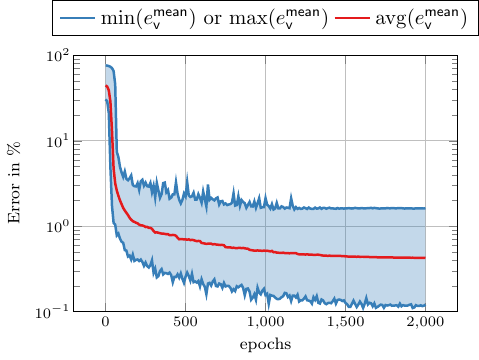}
        \caption{Validation set~$\setvalid$}
        \label{fig:valid_mean}
    \end{subfigure}
    
    \caption{Minimum, maximum, and average mean error~$\eoffmean$ of 
    all trained FDMNs for the training set~$\settrain$ (a) and the validation set~$\setvalid$ (b).}
    \label{fig:train}
\end{figure}
\foneme{
To measure the quality of fit, we define the mean error~$\eoffmean$ over a subset~$\mathcal{S}$ of the training data~$\setdata$ as 
\begin{equation}
    \eoffmean = \frac{1}{|\mathcal{S}|}\sum_{\left(\ffV_1^{\mysf{s}},\ffV_{\mysf{F}},\Veff^{\mysf{s}}\right)\in\,\mathcal{S}}
    \frac{\norm*{\DMNLIN(\ffV_1^{\mysf{s}},\ffV_{\mysf{F}},\vecalpha,\vecw) - \Veff^{\mysf{s}}}_1}{\norm*{\Veff^{\mysf{s}}}_1},
\end{equation}
where~$|\mathcal{S}|$ is the cardinality of the set~$\mathcal{S}$. 
Depending on the fiber orientation state~$\FOT$
and the initial guesses of the angles~$\vecalpha$ and the weights~$\vecw$, the quality of fit varies per DMN.
Therefore, for the whole training process of all 620 FDMNs, we visualize the
smallest mean error~$\min(\eoffmean)$, the largest mean error~$\max(\eoffmean)$, 
and the average mean error~$\mathrm{avg}(\eoffmean)$ for the training and validation
sets~$\settrain$ and~$\setvalid$ per epoch in Figure~\ref{fig:train}. For the training and the validation set, we observe that 
the three considered error measures drop rapidly at the beginning of the training,
and continue to improve as the training continues. As the learning rates are small for epochs larger than 1500, the errors
do not change significantly at the end of the training, and convergence is ensured. 
Evidently, the largest mean error~$\max(\eoffmeant)$ on the training set~$\settrain$ drops below 1\% at the end of the training process,
while the smallest mean error~$\min(\eoffmeant)$ falls below 0.3\%. 
This indicates a high quality of fit on the training set~$\settrain$ for all FDMNs.
Compared to the training set~$\settrain$, the spread between the largest mean error~$\max(\eoffmeanv)$ and the
smallest mean error~$\min(\eoffmeanv)$ for the validation set~$\setvalid$ is larger. 
However, the largest mean error~$\max(\eoffmeanv)$ 
stays below 2\% for all considered FDMNs and does not fluctuate by a large margin. 
Additionally, the average mean error~$\mathrm{avg}(\eoffmeanv)$ decreases continuously during training. 
Consequently, the prediction quality on both the 
training and validation set improve on average as the training progresses,
although the largest mean validation error~$\max(\eoffmeanv)$ does not improve significantly after 1250 training epochs.
Overall, this indicates that no pronounced overfitting to the training data set~$\settrain$ occurs.
}
\subsection{Online evaluation}

\foneme{To measure the online performance of an FDMN, we define the online error function~$\eon:\symdev\rightarrow\reals$ as
\begin{equation}
    \eon(\Deff) = \frac{\norm*{\fsigma^{\mysf{DMN}}(\Deff) - \fsigma^{\mysf{FFT}}(\Deff)}_2}{\norm*{\fsigma^{\mysf{FFT}}(\Deff)}_2},
\end{equation}
where~$\norm*{\cdot}_2$ denotes the $\ell^2$-norm of the components in Mandel notation,~$\Deff$ stands for 
the prescribed effective strain rate tensor, 
and~$\fsigma^{\mysf{DMN}}:\symdev\rightarrow\symdev$ and~$\fsigma^{\mysf{FFT}}:\symdev\rightarrow\symdev$
are the stress functions of the FDMN and the FFT-based homogenization, respectively.
We then choose the largest error over all load cases~$\eonmax$ and the mean error over 
all load cases~$\eonmean$
\begin{equation}
   \eonmax = \max_{\bm{\bar{D}} \in \bm{\bar{D}}_{\gdot}} \eon,\quad 
   \eonmean = \frac{1}{|\setD|}\sum_{\bm{\bar{D}}\in\bm{\bar{D}}_{\gdot}}\eon(\Deff),
\end{equation}
to study the performance of the different FDMNs. For each investigated orientation state~$\FOT$,
we identify the FDMN with the smallest maximum error~$\eonmax$ and visualize the associated 
errors~$\eonmean$ and~$\eonmax$ over the fiber orientation triangle, see Figures~\ref{fig:DOT_mean}
and~\ref{fig:DOT_max}
}
\begin{figure}[H]
    \centering 
    \begin{subfigure}{0.49\textwidth}
        \centering
        \includegraphics[trim={0 0 0 0},clip,width=\textwidth]{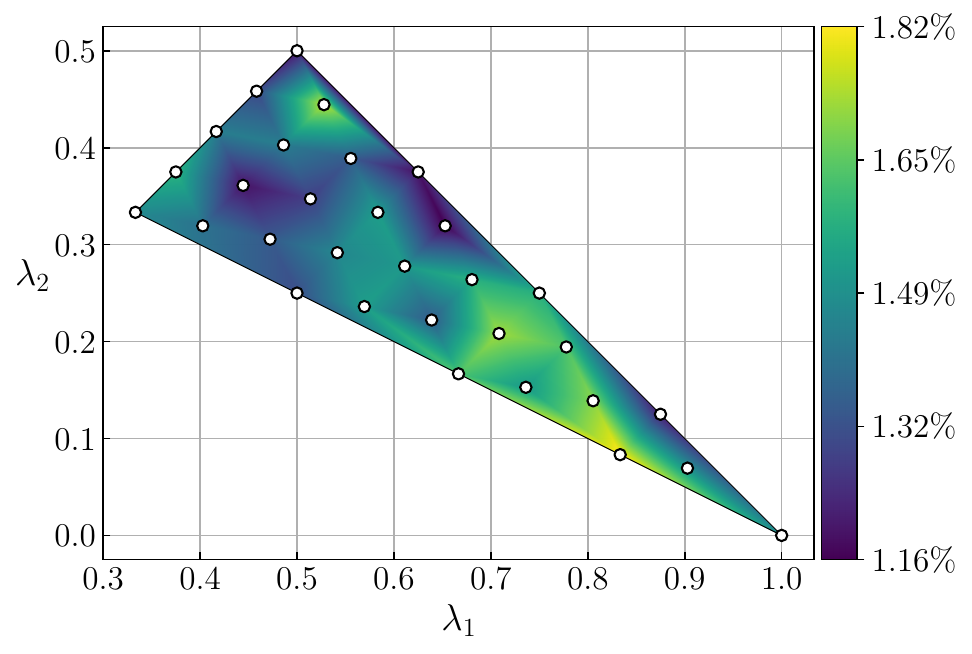}
        \caption{Mean error~$\eonmean$}
        \label{fig:DOT_mean}
    \end{subfigure}
    \begin{subfigure}{0.49\textwidth}
        \includegraphics[trim={0 0 0 0},clip,width=\textwidth]{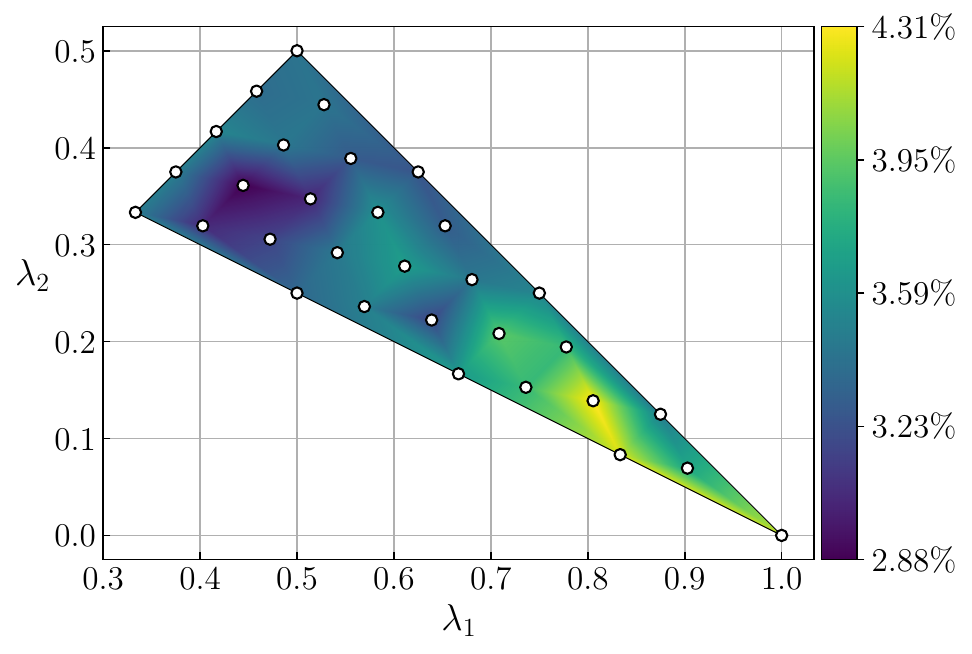}
        \caption{Largest error~$\eonmax$}
        \label{fig:DOT_max}
    \end{subfigure}
    
    \caption{Mean online error~$\eonmean$ (a) and largest online error~$\eonmax$ (b) for the best FDMNs over the
    fiber orientation triangle.}
    \label{fig:DOT}
\end{figure}
\foneme{
    The maximum error~$\eonmax$ of the best FDMNs range between 2.88\% for the orientation \linebreak 
    state~${\FOT=\T{(0.4444, 0.3611)}}$,
    and 4.31\% for the fiber orientation state~$\FOT=\T{(0.8055,0.1389)}$. Thus, for the investigated orientation states,
    the FDMNs yield sufficient prediction accuracy in terms of engineering requirements over a wide range of shear rates~$\gdot\in[10,10^5]\persec$. 
    Additionally, the relatively low maximum errors~$\eonmax$ show that with fewer samples
    than in previous studies for solid materials, \fthrtb{for example by Liu et al.~\cite{liu2019deep,liu2019exploring}
    or Gajek et al.~\cite{gajek2020micromechanics,gajek2021fe,gajek2022fe}},
    appropriately accurate FDMNs can be produced for suspensions of rigid fibers. 
    Generally, the maximum error~$\eonmax$ is larger for more strongly aligned oriented orientation states towards the lower right-hand side of the fiber orientation 
    triangle than for less strongly aligned oriented orientation states towards the \ftwome{upper and the} left-hand side.
    The mean error~$\eonmean$ ranges between 1.16\% for the orientation state~$\FOT=\T{(0.8334, 0.0833)}$,
    and 1.82\% for the orientation state~$\FOT=\T{(0.6528, 0.3194)}$, underlining the prediction accuracy of the FDMNs.

    To discuss one specific example, we take a more detailed look at the performance of the best FDMN at the
    fiber orientation state~$\FOT=\T{(0.8055,0.1389)}$ where the largest maximum error~$\eonmax=4.31\%$ occurs.
    Let the components of the strain rate tensor~$\Dload_{\mysf{i}}\in\symdev$, 
    in the standard basis~$\{\fe_i\}$ and in Mandel notation, 
    be given by the $\mysf{i}$-th column of the matrix~$\Dmatrix / \gdot$, see equation~\eqref{eq:Ds}.
    Then, the tensor~$\Dload_{\mysf{i}}$ represents a load with unit norm, and we refer to the
    tensor~$\Dload_{\mysf{i}}$ as load case. In the following, we discuss the observed stress responses and
    the prediction quality of the FDMN 
    for all considered shear rates~\eqref{eq:gammadots} and the six load cases 
    defined in equation~\eqref{eq:Ds}, see Figure~\ref{fig:0805}.
    Because of the strong alignment of the fibers in the coordinate direction~$\fe_1$, the stress norms are the largest for
    the load case~$\Dload_1$, which encodes incompressible elongational flow in the coordinate direction~$\fe_1$, see Figure~\ref{fig:0805_response}.
\begin{figure}[H]
    \centering 
    \begin{subfigure}{\textwidth}
        \centering
        \includegraphics[height=0.6cm]{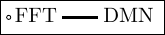}
        \includegraphics[height=0.6cm]{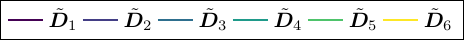}
    \end{subfigure}    
    \begin{subfigure}{0.49\textwidth}
        \centering
        \includegraphics[width=\textwidth]{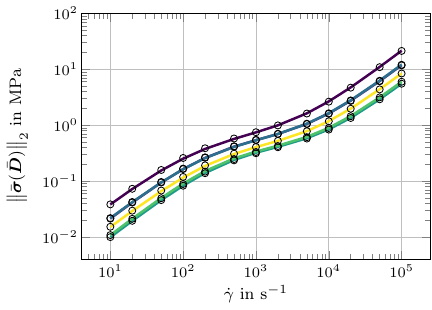}
        \caption{$\ell^2$-norm of the effective stress response~$\sigmaeff(\Deff)$}
        \label{fig:0805_response}
    \end{subfigure}
    \begin{subfigure}{0.49\textwidth}
        \includegraphics[width=\textwidth]{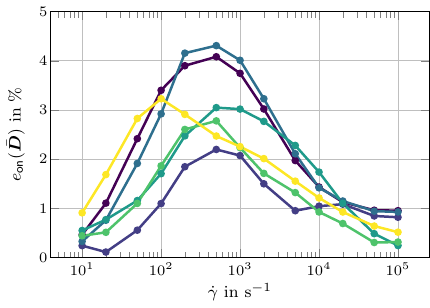}
        \caption{Online error~$\eon(\Deff)$}
        \label{fig:0805_error}
    \end{subfigure}
    
    \caption{$\ell^2$-norm of the effective stress response~$\sigmaeff(\Deff)$ over the
    shear rate~$\gdot\in[10,10^5]\persec$ at the orientation 
    state~${\FOT=(0.8055,0.1389)}$ as
    computed with FFT-based homogenization and the best identified FDMN (a). Online error~$\eon(\Deff)$ of the FDMN over 
    the shear rate~$\gdot\in[10,10^5]\persec$ (b).}
    \label{fig:0805}
\end{figure}The norms of the stress responses to elongational flow in the coordinate directions~$\fe_2$ and~$\fe_3$, i.e., the load cases~$\Dload_2$ and~$\Dload_3$,
are comparatively \ftwoms{small}. For the largest considered shear rate of~$\gdot=10^5\persec$, we observe a stress norm of 21~$\unit{MPa}$
for the load case~$\Dload_1$, as well as 11.8~$\unit{MPa}$ and 12.1~$\unit{MPa}$ for the load cases~$\Dload_2$ and~$\Dload_3$, respectively.
Because of the similar degrees of fiber alignment in the coordinate directions~$\fe_2$ and~$\fe_3$,
the stress norms associated with the load cases~$\Dload_2$ and~$\Dload_3$ differ only slightly.
Furthermore, the stress norms in response to shear in the load cases~$\Dload_4$ and~$\Dload_5$ are smaller than
to shear in the load cases~$\Dload_6$.
Again, this is caused by the strong alignment of the fibers in the coordinate direction~$\fe_1$,
which increases the flow resistance stronger in the~$\fe_1$-$\fe_2$-plane than in other shear planes~\cite{sterr2024machine}.
For the largest considered shear rate of~$\gdot=10^5\persec$, the stress norms are 5.5~$\unit{MPa}$, 6.0~$\unit{MPa}$, and 8.4~$\unit{MPa}$
for the load cases~$\Dload_4$,~$\Dload_5$, and~$\Dload_6$, respectively.
Overall, the predicted $\ell^2$-norm of the effective stress response~$\sigmaeff^{\mysf{DMN}}$ of the FDMN agrees closely with the
effective stress~$\sigmaeff^{\mysf{FFT}}$ computed via FFT-based homogenization for all investigated load 
directions~$\Dload_{\mysf{i}}, \, i \in\{1,...,6\}$, and shear rates~$\gdot\in\setgdot$, see Figure~\ref{fig:0805_response}.

Not only the~$\ell^2$-norm of the effective stress response is predicted well by the FDMN, but the direction as well.
This is evident from the relatively small online errors~$\eon(\Deff)$
shown in~\ref{fig:0805_error}. Depending on the load case, the error~$\eon$ increases up to a point
in the interval~$[10^2,10^3]\persec$, before decreasing again for higher shear rates. 
The largest error occurs for the elongational load case~$\Dload_3$ with 4.31\%, and is closely
followed by 4.08\% for the elongational load case~$\Dload_1$. For the shear load cases~$\Dload_4$,~$\Dload_5$, and~$\Dload_6$,
the largest errors are 3.05\%, 2.78\%, and 3.23\%, respectively, while the largest \ftwoms{error} for the elongational load
case~$\Dload_2$ is 2.2\%. The largest error for the shear load case~$\Dload_6$ occurs at a shear rate of~$\gdot=10^2\persec$,
whereas the largest errors for all other load cases occur at a shear rate of~$\gdot=5\cdot10^2\persec$.
This is directly tied to the Cross-type \fthrtb{constitutive equation}~\eqref{eq:crosseq}, which has two Newtonian
plateaus for low and high shear rates and a non-linear transition in between. 
Consequently, for large shear rates~$\gdot\rightarrow\infty$ the Matrix
behavior is Newtonian and the homogenization function of the microstructure approaches the linear homogenization function.
Since the FDMN approximates the non-linear homogenization function the microstructure to first order~\cite{gajek2020micromechanics},
the prediction quality depends on the degree to which the matrix behaves non-linearly.
This is in line with previous observations regarding the modelling of shear-thinning fiber 
suspensions~\cite{sterr2023homogenizing, sterr2024machine}.  
}
\foneme{
\subsection{\ftwome{Computational Speedup}}\label{subsec:computational_speedup}
\begin{table}[h!]
	\begin{center}
		\begin{tabular}{l c c c}
			\hline
			& Sampling total &  Sampling per orientation (mean) & Training \\
			\hline
			CPU Hours & 24415 & 788 & 0.08 \\
			Wall-clock time (h) & 305  & 9.8 & 0.08  \\
			\hline
		\end{tabular}
	\end{center}
	\caption{CPU hours and \ftwoms{w}all-clock time for the training of one FDMN
    and the sampling of effective viscosities~$\Veffsample$ using FFT-based homogenization.}
	\label{tab:sampling_training}
\end{table}
\begin{table}[h!]
	\begin{center}
		\begin{tabular}{l c c c}
			\hline
			& FFT (h) & DMN (ms) & Speedup\\
			\hline
			Min   & 2.08 & 600 & 11785\\
			Mean  & 2.61 & 630 & 14828\\
			Max   & 3.07 & 650 & 17225\\
			\hline
		\end{tabular}
	\end{center}
	\caption{Speedups and \ftwoms{w}all-clock times for the evaluation of a single load case
    using FFT-based homogenization compared to an FDMN.}
	\label{tab:online}
\end{table}
For sampling the linear homogenization functions of the 31 considered fiber orientation states~$\FOT\in\setFOT$,
we used a workstation with two AMD EPYC 7552 48-Core processors and 1024 GB DRAM. We \ftwoms{relied on} a single thread
per FFT-based computation and ran 80 computations in parallel. In total, the computations took 24415 CPU hours
and a \ftwoms{w}all-clock time of 305 hours, see Table~\ref{tab:sampling_training}.
Averaging over all 31 considered orientations~$\FOT$, this leads to 788 CPU Hours and~$9.8\unit{h}$ \ftwoms{w}all-clock time per orientation.
For the training and the online evaluation of the FDMNs, as well as the non-linear
FFT-based computations, we used a workstation
with two AMD EPYC 9534 64-Core Processors and 1024 GB DRAM, and ran all computations on a single thread. 
The \ftwoms{w}all-clock time to evaluate all six load cases~$\Dload_{\mysf{i}},\, i=1,...\fthrtb{,}6$, and thirteen shear rates~$\gdot\in\setgdot$
in series for a single fiber orientation state~$\FOT$ is listed in Table~\ref{tab:online}. The \ftwoms{w}all-clock
times for the FFT-based computations range from~$2.08\unit{h}$ to~$3.07\unit{h}$ with the mean \ftwoms{w}all-clock time over all
considered orientation states~$\FOT$ being~$2.61\unit{h}$. In contrast, the online evaluation of the FDMNs
takes between~$600\unit{ms}$ and~$630\unit{ms}$, with a mean of~$650\unit{ms}$. Consequently, the speedup factors 
range between 11785 and 17225 with a mean of 14828. This considerable speedup is achieved by investing computational
resources into the sampling of the linear homogenization functions of the microstructures. It is straightforward
to judge whether this investment is sensible for the considered setup. For one microstructure the generation of training data
and the training of a single FDMN takes~$9.88\unit{h}$ \ftwoms{w}all-clock time on average, while the non-linear FFT-based computations
take~$2.61\unit{h}$ \ftwoms{w}all-clock time on average. Therefore, assuming the non-linear computations run on \ftwoms{one}
thread as in the presented setup, it would take \ftwoms{four} non-linear computation to offset the initial investment into an FDMN on average. Given that component scale simulations
routinely require thousands or millions of microscale computations~\cite{gajek2021fe,gajek2022fe}, the sampling and training
effort is offset easily in engineering problems. 
}

\ftwoms{
\subsection{Comparison with machine learning aided analytical models}\label{subsec:FDMNs_vs_models}
In this section, we compare the FDMN based approach to predicting
the effective behavior of fiber suspensions with another machine learning
approach suggested by Sterr et al.~\cite{sterr2024machine}. 
\fthrms{The authors} presented four different analytical
\fthrtb{constitutive equation}s for the effective viscosity of shear-thinning fiber suspensions,
and identified the model parameters using supervised machine learning.
The model parameters were learned from simulation data obtained with FFT-based computational techniques
for the same material as considered in this article:
\fthrms{r}igid fibers suspended in a Cross-type matrix material with parameters as shown in Table~\ref{tab:crossparams}.
However, the FDMN approach and the analytical approach follow very different paradigms.
The FDMNs are trained to approximate the non-linear homogenization function of the microstructure
by learning the respective \fthrms{\textit{linear}} homogenization function without prior knowledge of the actual
\fthrtb{constitutive equation}s. In contrast, the parameters of the analytical models
were learned from the \fthrms{\textit{non-linear}} stress response of the suspension,
and the analytical models incorporate knowledge about the expected material behavior. 
Even though the two approaches follow different strategies, they achieve comparable prediction accuracy.
For the load cases~$\setD$ and fiber orientation states~$\FOT$ considered in this article, the largest validation error
of the FDMNs is~$4.31\%$ and occurs for the fiber orientation state~$\FOT=\T{(0.8055, 0.1389)}$.
For the same load cases~$\setD$ and the same set of fiber orientation states~$\FOT$,
three of the four analytical models by Sterr et al.~\cite{sterr2024machine}
achieve a maximum prediction error of~$5.00\%$, which occurs for the fiber orientation state~${\FOT = \T{(0.8334, 0.0833)}}$.
The fourth model did not compare favorably with the other three, because the built-in assumptions of stress-strain rate
superposition and orientation averaging did not hold well for the considered suspension~\cite{sterr2024machine}.
Consequently, the FDMNs achieved a \fthrms{slightly} higher prediction accuracy for the type of fiber suspension considered in this article.
Also, within the constraints of a first order approximation, FDMNs generalize to different \fthrtb{constitutive equation}s while the 
presented analytical models do not, reducing the required modelling effort for FDMNs.
However, the FDMNs utilized in this article use 1150 free parameters,
whereas the analytical models use between 11 and 49 parameters.

In terms of computational cost, generating the training data and training an FDMN for a single microstructure took~$9.88\unit{h}$ wall-clock time on average, see Table~\ref{tab:sampling_training}.
The parameters of the analytical models for a single microstructure were identified based on six non-linear FFT-based simulations,
which took~$2.61\unit{h}$ wall-clock time on average. This leads to an average total wall-clock time of~$15.70\unit{h}$
if all required simulations are run in series on one thread, and~$\unit[0.04]{h}$ are allocated to identify the model parameters.
However, with the considered setup, the non-linear FFT-based simulations could be run in parallel,
such that the analytical model parameters can be obtained more quickly than an FDMN. 
In summary, FDMNs offer a higher degree of accuracy and flexibility regarding the considered \fthrtb{constitutive equation}s, but require
more computational resources and a larger amount of parameters. \fthrms{Furthermore,
FDMNs are inherently thermodynamically consistent~\cite[§3.1]{gajek2020micromechanics},
and inherit stress-strain rate monotonicity from their phases~\cite[App. C]{gajek2020micromechanics}.
In contrast, both thermodynamic consistency and stress-strain rate monotonicity must be ensured manually in constitutive models.
}
}

\section{Conclusions}
\ftwome{
In this article, we extended the direct DMN architecture to the Flexible DMN (FDMN) architecture for the treatment of 
fiber suspensions with infinite material contrast and shear-thinning matrix behavior.
To do so, we derived linear homogenization functions for two-phase layered emulsions that are governed by Stokes flow
and consist of linearly viscous phases. More specifically, we utilized 
results by Kabel et al.~\cite{kabel2016model} and Milton~\cite{milton2022theory}
on the homogenization of laminates to derive closed form analytical expressions
for the effective properties of such layered emulsions.
Because rank-one layered materials are ill-suited as DMN building blocks in case of infinite material
contrast, we investigated the \fthrms{effective properties} of coated layered materials (CLMs). We leveraged the
linear homogenization functions of rank-one layered materials and investigated
under which conditions the effective behavior of CLMs is 
non-singular if the core material is \fthrms{rigid}. The conditions
depend on the physical constraints of the employed materials, and involve the rank of the CLM
and the relative orientation of the layering directions. In the relevant case for incompressible fiber suspensions,
a CLM consists of an incompressible coating material and a rigid core material. Then, the 
effective \fthrms{viscosity} of a rank-$3$ CLM \fthrms{is} non-singular if the three layering directions are mutually non-orthogonal
and mutually non-collinear. 

Using the derived homogenization functions for layered materials and CLMs,
we extended the (direct) DMN architecture to the FDMN architecture
by replacing the lowest layer of rank-one laminates in a (direct) DMN with non-singular CLMs, and the other
rank-one laminates with rank-one layered materials capable of treating fluids. 
For the offline training of an FDMN, we presented a strategy where the relative angles of the CLM
layering directions are fixed to reduce the amount of free parameters and guarantee non-singular CLMs.
Furthermore, we modified the online evaluation strategy of direct DMNs to account for incompressible phases.
We leveraged the FDMN architecture to predict the non-linear effective behavior of fiber suspensions
with a Cross-type matrix material. Compared to direct numerical simulations with FFT-based computational techniques,
the FDMNs achieved validation errors below~$4.31\%$ for a variety of 31 fiber orientation states, six 
different load cases, and a wide range of shear rates relevant to engineering processes. 
If the time required to generate the training data and train the FDMNs is not considered,
the FDMNs achieved an average speed up factor of 14828 as compared to FFT-based simulations.
Additionally, FDMNs achieve higher accuracy than another machine learning based approach by Sterr et al.~\cite{sterr2024machine}
for the same composite material. However, this \fthrms{competitive} accuracy improvement comes at 
\fthrms{the cost of an increased computational effort to obtain} the training data for the FDMNs.

In future work, the application and extensions of the FDMN architecture to problems involving more
than two phases, where multiple phases may be singular, could be explored. This could enable the
FDMN based treatment of three-phase systems containing a fluid, fibers, and gas,
which occur in engineering processes like injection molding~\cite{shuler1994rheological}
and flotation de-inking~\cite{cui2007flow}. Furthermore, problems involving solids
with defects could be explored \fthrme{a}s well.
To account for the locally varying microstructure in component scale simulations,
fiber orientation interpolation schemes~\cite{huang2022microstructure,gajek2021fe,dey2024effectiveness}
could be combined with the FDMN architecture to increase versatility, and enable 
new concurrent two-scale simulations involving defects and rigid inclusions.

}

\section*{Declaration of competing interest}
The authors declare that they have no known competing financial
interests or personal relationships that could have appeared to
influence the work reported in this paper.

\section*{Data availability}
The data that support the findings of this study are available from the corresponding author upon reasonable request.

\section*{Acknowledgements}
The research documented in this manuscript was funded by the Deutsche Forschungsgemeinschaft (DFG, German Research Foundation), project number 255730231, within the International Research Training Group “Integrated engineering of continuous-discontinuous long fiber-reinforced polymer structures“ (GRK 2078/2). The support by the German Research Foundation (DFG) is gratefully acknowledged.

\section*{Author contributions}
\ftwome{The present study was conceptualized by B. Sterr, M. Schneider and T. Böhlke. The presented
FDMN architecture was derived by B. Sterr, S. Gajek, M. Schneider, and T. Böhlke. B. Sterr, S. Gajek, and M. Schneider
implemented and validated the software. B. Sterr performed the simulations, analyzed and visualized the data,
and drafted the manuscript. The original manuscript draft was extensively reviewed and edited by B. Sterr, A. Hrymak,
M. Schneider, and T. Böhlke. Resources were provided by M. Schneider and T. Böhlke. The research project was supervised
by A. Hrymak, M. Schneider and \mbox{T. Böhlke}.
}
\newpage
\setcounter{subsection}{0}
\renewcommand{\thesubsection}{\Alph{subsection}}
\section*{Appendix}
\ftwoms{\subsection{Coated layered materials with singular core}\label{sec:regularityconditions}
\subsubsection{Singularity condition for coated layered materials with singular core}}
\foneme{In the particular cases that the core material~$m_1$ of a CLM is rigid or represents a void, 
the dual or the primal material properties~$\ffK_1$ or~$\ffM_1$ approach zero, respectively.
Thus, more generally, we let the material tensor~$\ffA_1$ approach zero, and rewrite equation~\eqref{eq:rankthree_milton} as
\begin{equation}\label{eq:a2limitequation}
    \lim_{\ffA_1\rightarrow\bm{0}}\Aeff = \ffA_2 \left( \ffI + (1-f_2)\left(-\ffI + f_2 \sum_{\mysf{r}=1}^{R}\csfr\ftwoms{\GammaAtwo}(\fnr)\ffA_2\right)^{-1}\right).
\end{equation}
\ftwoms{In the following, we study under which conditions the effective material properties~$\Aeff$ of a rank-$R$
CLM are singular.}
The effective material properties~$\Aeff$ are singular \fthrms{precisely if}
there exists an effective tensor~$\geff\in\{\sym,\symdev\}$, such that
\begin{equation}\label{eq:Agzero}
    \Aeff[\geff] = \fzero\fthrms{.}
\end{equation}
\fthrms{Using the definition~\eqref{eq:a2limitequation},
equation~\eqref{eq:Agzero} may be equivalently rewritten in the form}
\begin{equation}\label{eq:evproblem}
    \left(\sum_{\mysf{r\ftwome{=1}}}^{R}\csfr\ftwoms{\GammaAtwo}(\fnr)\ffA_2\right)[\geff] = \geff.
\end{equation}
Equation~\fthrms{\eqref{eq:evproblem}} is satisfied \fthrms{precisely if}~$\geff$ lies in the 
intersection of all subspaces~$\spaceAstrainsfr$, such that
\begin{equation}\label{eq:inspace1}
    \geff \in \bigcap_{\mysf{r}=1}^{R} \ftwome{\spaceAstrainsfr},
\end{equation}
or, equivalently,
\begin{equation}\label{eq:inspace2}
    \fthrms{\GammaA(\fn_{r})[\geff] = \geff, \quad \forall r \in \{1,...,R\}.}
\end{equation}
We prove this as follows.
First, if~$\geff$ lies in the section of all subspaces such that~\fthrms{\eqref{eq:inspace1}} is satisfied,
it follows \fthrms{from equation~\eqref{eq:inspace2} and the definition of the operator~$\GammaAtwo(\fnr)$~\eqref{eq:Gammadef_Atwo}} that
\begin{equation}
    \left(\sum_{\mysf{r\ftwome{=1}}}^{R}\csfr\ftwoms{\GammaAtwo}(\fnr)\ffA_2\right)[\geff] = 
    \left(\sum_{\mysf{r\ftwome{=1}}}^{R}\csfr\ftwoms{\GammaAtwo}(\fnr)\ffA_2\GammaA(\fn_r)[\geff]\right) = 
    \sum_{\mysf{r}\ftwome{=1}}^{R}\csfr\geff = \geff,
\end{equation}
and hence equation~\eqref{eq:evproblem} is satisfied.
Second we wish to prove that the converse is also true, i.e., if equation~\eqref{eq:evproblem} is satisfied, 
equation~\eqref{eq:inspace1} is satisfied as well. 
\ftwoms{Let~$\ffT_{\ffA_2}$
denote a convex combination of the projectors~$\GammaAtwo(\fnr)\ffA_2$, i.e.,
\begin{equation}
    \ffT_{\ffA_2} = \left(\sum_{\mysf{r\ftwome{=1}}}^{R}\csfr\ftwoms{\GammaAtwo}(\fnr)\ffA_2\right)\fthrms{,}
\end{equation}
\fthrms{with
\begin{equation}
    c_r > 0 \quad\text{and}\quad \sum_{r=1}^R c_r = 1.
\end{equation}}Then,
\fthrms{fixed points of the} operator~$\ffT_{\ffA_2}$ \fthrms{lie in the} set~\cite[Lemma 1.4]{reich1983limit}
\begin{equation}
    \mathcal{E} = \bigcap_{\mysf{r}=1}^{R} \ftwome{\spaceAstrainsfr}.
\end{equation}
Therefore, any eigenvector of the operator~$\ffT_{\ffA_2}$
corresponding to the eigenvalue one must lie in the intersection of the subspaces~$\spaceAstrainsfr$,
which we denote with~$\mathcal{E}$. In particular, if
equation~\eqref{eq:evproblem} is satisfied, the element~$\geff$ lies in $\mathcal{E}$, \fthrms{i.e.,
\begin{equation}\label{eq:liesinspace}
    \ffT_{\ffA_2}[\geff] = \geff,\quad\implies\quad \geff \in\mathcal{E}
\end{equation}
}and equation~\eqref{eq:inspace1} is satisfied as well.
}

\fthrms{With the result~\eqref{eq:liesinspace} at hand, we observe that a necessary and sufficient condition
for the regularity of a rank-R CLM is that the space~$\mathcal{E}$ is trivial, i.e.,}
\begin{equation}\label{eq:non-singularitycondition}
    \mathcal{E} = \{\fzero\}.
\end{equation}
We refer to equation~\eqref{eq:non-singularitycondition} as \textit{non-singularity condition} in the following.
In the context of linear elastic and linearly viscous materials, we
derived the non-singularity condition~\eqref{eq:non-singularitycondition} for 
the \ftwoms{spaces~$\spaceAstrainsfr$, such that
\begin{equation}
    \spaceAstrainsfr \subset \sym, \quad \forall \mysf{r}\in\{1,...,R\} 
    \quad\text{or}\quad
    \spaceAstrainsfr \subset \symdev, \quad \forall \mysf{r}\in\{1,...,R\}.
\end{equation}}However,
the presented proof extends to other problems that can be formulated in
the form of equation~\eqref{eq:a2limitequation}, such as thermoelastic or
piecoelectricity problems~\cite[§9]{milton2022theory}.
With the goal of using CLMs as building blocks in a DMN architecture, we are interested
\fthrme{in whether} there are particular choices of the layering directions~$\fnr$ and the 
rank~$R$ of the CLM for which the effective properties of a CLM are always 
non-singular, i.e., the non-singularity condition~\eqref{eq:non-singularitycondition}
is satisfied. 
\fthrms{For incompressible suspensions of rigid fibers, we study this question for CLMs with
a \fthrms{rigid} core material~$m_1$ and an incompressible coating material~$m_2$.}
\subsubsection{Coated layered materials with incompressible coating and rigid core}\label{subsec:CLMs_ico_rigid}
For applications involving rigid inclusions in incompressible media,
we consider CLMs with an incompressible coating material~$m_2$ and a rigid core material~$m_1$.
Therefore, the spaces~$\spaceAstrainsfr\subset\symdev$
have the form
\begin{equation}\label{eq:space_ico_rigid}
    \spaceAstrainsfr = \{\fB\in\symdev \quad|\quad  
    \fB\cdot(\fnr\otimessym\fa) = 0,\quad\forall \fa\in\reals^3\},
\end{equation}
with the dimensions of the spaces~${\spaceAstrainsfr}$
and~$\symdev$ given by
\begin{equation}
    \dim(\spaceAstrainsfr)=3, \quad\text{and}\quad \dim(\symdev)=5.
\end{equation}
Because the dimension~$\dim(\spaceAstrain_1+\spaceAstrain_2)$ is bounded by
the dimension of the space~$\symdev$, such that
\begin{equation}
    \dim(\spaceAstrain_1+\spaceAstrain_2) \leq 5,
\end{equation}
\fthrms{it follows from the dimension of the intersection of the two spaces~$\spaceAstrain_1$ and~$\spaceAstrain_2$
\begin{equation}
    \dim(\spaceAstrain_1\cap\spaceAstrain_2) = \dim(\spaceAstrain_1) + \dim(\spaceAstrain_2) - \dim(\spaceAstrain_1+\spaceAstrain_2),
\end{equation}
that}
\begin{equation}
    \dim(\spaceAstrain_1\cap\spaceAstrain_2) \geq 3 + 3 - 5 = 1.
\end{equation}
In other words, the intersection of the spaces~$\spaceAstrain_1$ and~$\spaceAstrain_2$ is at least one dimensional.
Consequently, \fthrms{two layering directions are not sufficient}
to satisfy the condition~\eqref{eq:non-singularitycondition}
Besides the required number of layerings,
we are also interested \fthrme{in} whether there are restrictions on the angles between the layering directions.
\fthrms{Suppose that one layering direction is orthogonal to the other two.
We discuss the case that the direction $\fn_3$ is orthogonal to the normals $\fn_1$ and $\fn_2$.
The other cases work similarly via permuting the indices.
Then it holds that}
\begin{equation}
    \fn_3\cdot\fn_1 = 0,\quad\text{and}\quad \fn_3\cdot\fn_2 = 0,
\end{equation}
\fthrms{and} the intersection~$\spaceAstrain_1\cap\spaceAstrain_2\cap\spaceAstrain_3$
does not only contain the zero element~$\fzero$. \fthrms{Hence,} the condition~\eqref{eq:non-singularitycondition}
is violated. To show this, we consider the alternative description of the spaces~$\spaceAstrainsfr$
\begin{equation}
    \spaceAstrainsfr = \left\{\fB \in \symdev \quad|\quad \fB\fnr = \alpha\fthrms{\fnr}\quad\text{\ftwoms{for some}}\quad \alpha\in\fthrms{\reals}\right\}
\end{equation}
which is equivalent to equation~\eqref{eq:space_ico_rigid} and characterizes the space\ftwome{~$\spaceAstrainsfr$} 
via all tensors~$\fB\in\spaceAstrainsfr$ that have the layering direction~$\fnr$ as 
\fthrms{an eigenvector}. Then the tensor
\begin{equation}
    \fB = \fn_1\otimes\fn_1 + \fm_2\otimes\fm_2 - 2\fn_3\otimes\fn_3,
\end{equation}
where the layering direction~\fthrms{$\fm_2$} is constructed by orthogonalization, such that
\begin{equation}
    \fm_2 = \fn_2 - (\fn_1\cdot\fn_2)\fn_1,
\end{equation}
is an element of the intersection~$\spaceAstrain_1\cap\spaceAstrain_2\cap\spaceAstrain_3$.
This is true because it holds that
\begin{equation}
    \tr(\fB) = 0,\quad\fB\fn_1 = \fn_1,\quad \fB\fn_2 = \fn_2,\quad \fB\fn_3 =\fn_3,
\end{equation}
and therefore condition~\eqref{eq:non-singularitycondition} is violated.
In other words, the effective properties~$\Aeff$ of a rank-$3$ CLM with 
incompressible coating and rigid core, where one layering direction~$\fn_3$
is orthogonal to the other two layering directions~$\fn_1$ and~$\fn_2$, is always singular.
\fthrms{Particularly, if all three layering directions are mutually orthogonal, the effective properties~$\Aeff$ of 
a rank-$3$ CLM are singular.}

\fthrms{In contrast}, let the three layering directions be mutually non-orthogonal and mutually non-collinear, i.e.,
\begin{equation}\label{eq:mutuallynon}
    0 < |\fn_1\cdot\fn_2| < 1 ,\quad 0<|\fn_1\cdot\fn_3|<1 ,\quad 0<|\fn_2\cdot\fn_3|<1.
\end{equation}
\fthrms{We consider an element~$\fB$ in the intersection~$\spaceAstrain_1\cap\spaceAstrain_2\cap\spaceAstrain_3$
and wish to show that the tensor~$\fB$ vanishes,
such that the non-singularity condition~\eqref{eq:non-singularitycondition} is satisfied.}
Because the tensor~$\fB$ is symmetric, its eigenvectors 
corresponding to distinct eigenvalues are orthogonal. 
\fthrms{Thus, if the eigenvalues were distinct, the 
layering directions would need to be orthogonal. However, this contradicts our assumption~\eqref{eq:mutuallynon}.}
Therefore,
the tensor~$\fB$ has a single eigenvalue~$\alpha$ with the corresponding eigenspace~$\reals^3$,
\fthrms{i.e., the tensor~$\fB$ attains the form}
\begin{equation}
    \fB = \alpha\fI, \quad \alpha\in\reals.
\end{equation}
Additionally, because the trace of the tensor~$\fB$ vanishes, i.e., ~$\tr(\fB)=0$, it follows that the eigenvalue~$\alpha$ is zero.
\foneme{Consequently, \fthrms{the tensor~$\fB$ must vanish, }the non-singularity condition~\eqref{eq:non-singularitycondition} is satisfied, and the effective
properties~$\Aeff$ of a rank-$3$ CLM with an incompressible coating material~$m_2$ and a rigid core material~$m_1$ are non-singular,
if the three layering directions are mutually non-orthogonal and non-collinear.}
}

\bibliographystyle{ieeetr}
{\footnotesize\bibliography{literature}}

\end{document}